\documentclass[sn-mathphys]{sn-jnl_no-header}

\usepackage{siunitx}

\usepackage{bm}

\usepackage{xcolor}

\jyear{2021}%

\theoremstyle{thmstyleone}%
\theoremstyle{thmstyletwo}%

\theoremstyle{thmstylethree}%

\newcommand{\covid}{\textsc{covid}-19}
\newcommand{\sars}{\textsc{sars}-\textsc{c}o\textsc{v}-2}

\usepackage{multicol}

\begin{document}

\title[]{Mobility and the spatial spread of \sars{} in Belgium}


\author*[1]{\fnm{Michiel} \sur{Rollier}}\email{michiel.rollier@ugent.be}
\equalcont{These authors contributed equally to this work.}

\author[1,3]{\fnm{Gisele H. B.} \sur{Miranda}}
\equalcont{These authors contributed equally to this work.}

\author[1,2]{\fnm{Jenna} \sur{Vergeynst}}
\equalcont{These authors contributed equally to this work.}

\author[1]{\fnm{Joris}
\sur{Meys}}

\author[2]{\fnm{Tijs W.}
\sur{Alleman}}


\author[5]{\fnm{the Belgian Collaborative Group}\sur{ on \covid{} Hospital Surveillance}}


\author[1]{\fnm{Jan M.}
\sur{Baetens}}

\affil*[1]{\orgdiv{KERMIT}, \orgname{Department of Data Analysis and Mathematical Modelling, Ghent University}, \orgaddress{\street{Coupure Links 653}, \city{Ghent}, \postcode{9000}, \country{Belgium}}}

\affil[2]{\orgdiv{BIOMATH}, \orgname{Department of Data Analysis and Mathematical Modelling, Ghent University}, \orgaddress{\street{Coupure Links 653}, \city{Ghent}, \postcode{9000}, \country{Belgium}}}

\affil[3]{\orgdiv{Division of Computational Science and Technology}, \orgname{KTH Royal Institute of Technology}, \orgaddress{\street{Tomtebodavägen 23A}, \city{Solna}, \postcode{17165}, \country{Sweden}}}

\affil[4]{\orgdiv{Department of Environment}, \orgname{Ghent University}, \orgaddress{\street{Coupure links 653}, \city{Ghent}, \postcode{9000}, \country{Belgium}}}

\affil[5]{\orgdiv{Department of Epidemiology and Public Health}, \orgname{Sciensano}, \orgaddress{, \city{Brussels}, \postcode{1050 Brussels}, \country{Belgium}}}


\abstract{We analyse and mutually compare time series of \covid{}-related data and mobility data  across Belgium's 43 arrondissements (NUTS 3).
In this way, we reach three conclusions.
First, we could detect a decrease in mobility during high-incidence stages of the pandemic. This is expressed as a significant change in the average amount of time spent outside one's home arrondissement, investigated over five distinct periods, and in more detail using an inter-arrondissement ``connectivity index'' (CI).
Second, we analyse spatio-temporal \covid{}-related hospitalisation time series, after smoothing them  using a generalise additive mixed model (GAMM). We confirm that some arrondissements are ahead of others and morphologically dissimilar to others, in terms of epidemiological progression. The tools used to quantify this are time-lagged cross-correlation (TLCC) and dynamic time warping (DTW), respectively.
Third, we demonstrate that an arrondissement's CI with one of the three identified first-outbreak arrondissements is correlated to a significant local excess mortality some five to six weeks after the first outbreak. More generally, we couple results leading to the first and second conclusion, in order to demonstrate an overall correlation between CI values on the one hand, and TLCC and DTW values on the other. We conclude that there is a strong correlation between physical movement of people and viral spread in the early stage of the \sars{} epidemic in Belgium, though its strength weakens as the virus  spreads}.




\maketitle

\section{Introduction}\label{sec:introduction}

\begin{figure}[h]
    \centering
    \includegraphics[width=.8\linewidth]{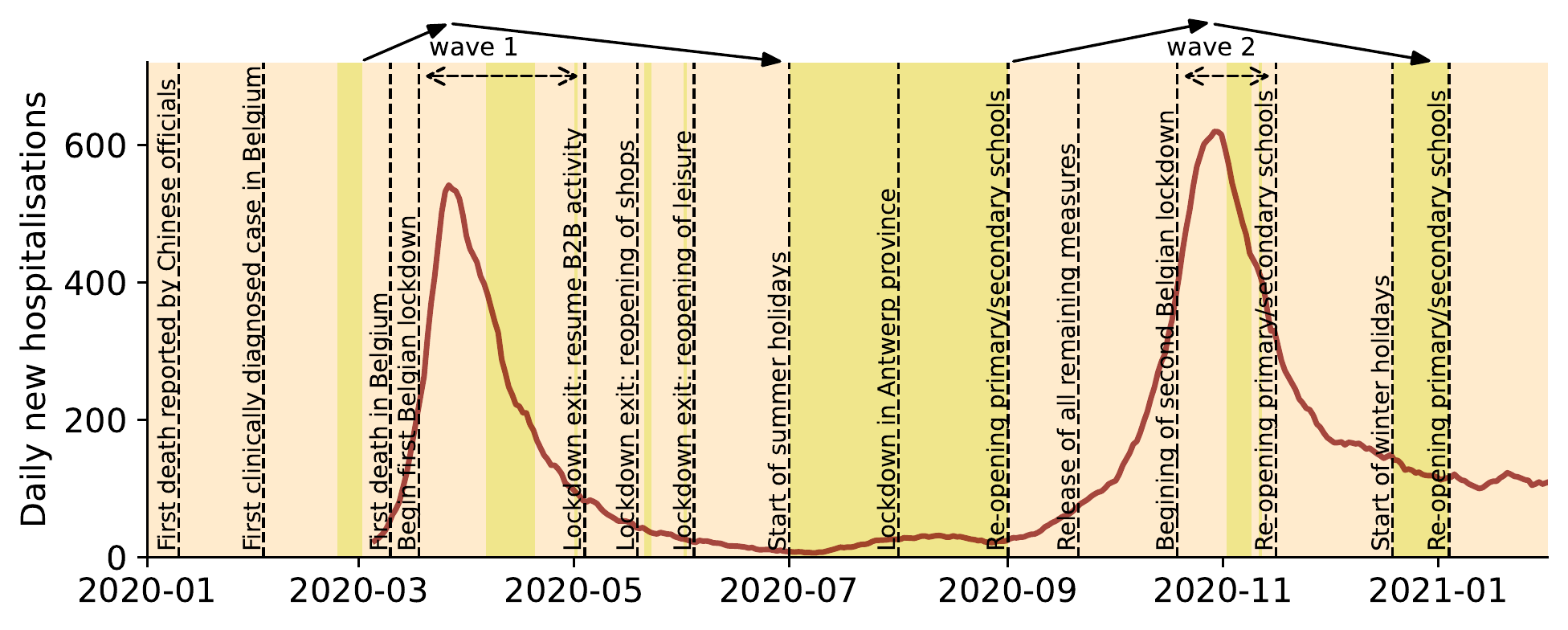}
    \caption{2020 Timeline  with the indication  of major \covid{}-related events (dashed vertical lines). The maroon-coloured graph represents the 7-day centralised moving average of daily new hospitalisations, aggregated for the whole of Belgium, obtained from Sciensano \citep{Sciensano2020}. The dashed horizontal arrows denote periods with lockdown measures. The solid sloped arrows demarcate the periods defined as wave 1 and wave 2 in our analysis (see Section \ref{subsec:demarcations}). Khaki-coloured areas represent conventional school holidays.}
    \label{fig:timeline_2020}
\end{figure}

\covid{} is a respiratory disease caused and spread by \sars{}, an infectious coronavirus.  The first confirmed case in Belgium was a repatriated person from Wuhan \citep{FPS1}, who tested positive on February 4, 2020 but did not spread the disease further (see Fig.~\ref{fig:timeline_2020} for a chronological overview). Returning vacationers from affected regions in Italy led to additional importations during the half-term holidays \citep{FPS2}, and by early March 2020, disease transmission in Belgium was confirmed. On March 13, 2020, the Belgian government imposed the first measures to control the virus spread. 
On March 18, 2020, the measures were tightened to a lockdown. 
Two days later the Belgian borders were closed for non-essential travel, meaning that from that moment on cross-border mobility was severely restrained. Restrictions were gradually eased in May, June and July 2020. Upon the emergence of a second \covid{} wave in September-October, measures were again restricted on October 19, 2020.\\


To inform policymakers about the forecasted evolution of the pandemic and the projected effects of containment measures, we developed a  compartmental \covid{} metapopulation model operational at the national scale \citep{Alleman2021} that tracks the number of individuals in 10 different epidemiological and clinical stages and 9 different age groups. It was developed in parallel with other researchers with slightly varying approaches \citep{Abrams2020,Franco2020}. The collective results of these efforts  were finally bundled to obtain ensemble forecasts \citep{RESTORE8}.\\

The aforementioned models work at the level of the entire Belgian population and hence do not capture the spatially heterogeneous nature of  disease spread. However, spatially explicit phenomena such as human mobility have been shown to play an important role in the emergence of spatial disease dynamics in general \citep{Changruenngam2020, Merler2010, Wesolowski2012, Wesolowski2015}.
For what concerns \covid{} in particular, Refs. \cite{du2020, jia2020} have shown the importance of population flow on the dynamics of the emerging pandemic in China.
On a smaller geographical scale and in a western context, Iacus et al. \cite{Iacus2020} demonstrated that the initial \covid{} spread in France and Italy can be explained to a large extent by mobility between departments.
A model that aims to correctly describes the spread of \sars{} in large and diverse nations, therefore benefits from including a notion of population dynamics. Refs. \cite{Arenas2020, Costa2020, Roques2020, sartorius2021} have developed such models for resp. Spain, Brazil, France and the United Kingdom.\\

Adding a spatial component to our model \cite{Alleman2021} as well is both \textit{feasible} and \textit{informative}, as the relevant data are available and there is a demand for local insights and projections. Belgium, however, one would not typically classify as a large and diverse nation; with a road density of more than 500 km per 100 km$^2$ \citep{Knoema2011} and a 2020 population density of 374 inhabitants per km$^2$  \citep{STATBEL2020}, Belgium is one of the most connected and densely populated countries in the world. It is therefore not clear whether the relations between human mobility and \sars{} spread observed in large countries, are also convincingly expressed for Belgium. In other words, it is not clear whether the inclusion of human mobility into a spatially explicit \sars{} model is also \textit{necessary}. The final objective of this study is to address that question.\\

The question is tackled by quantitatively addressing three sub-questions, all dealing with Belgian mobility and epidemiological time series in the year 2020.
We first focus on mobility time series only and demonstrate that
the average amount of time people spend outside their home arrondissement strongly declines during periods of elevated \covid{} hospitalisation. This serves as a sanity check, and is primarily addressed to define the ``connectivity index''.
Second, we focus on \covid{}-related time series only, quantitatively discerning spatiotemporal structure in the spread of the disease. We do so by investigating, in terms of local \covid{} hospitalisation time series, which arrondissements are ``running ahead'' of others, and which arrondissement pairs showed a morphologically similar \covid{} evolution. We do so by applying time-lagged cross correlation and dynamic time warping, respectively.
Third, these separate approaches are combined.
In particular, we show that the strength of the connection to an early-outbreak arrondissement appears to predict significant excess mortality five to six weeks after the onset of the pandemic.

In general, our analysis indicates quantitatively that \covid{}-related time series of strongly connected arrondissements are on average more synchronised and morphologically similar than those of poorly connected arrondissements.
This suggests that a comprehensive \covid{} model for Belgium or similarly small countries may benefit from taking into account (fluctuating) mobility patterns. With such a model, the impact of regulating mobility during the \textit{early} stages of a pandemic may be of particular interest.\\

Below, in Section \ref{sec:data-collection}, we first sketch the geographical situation in more detail and discuss the data used in our analysis. Section \ref{sec:methods} present the methods for summarising and smoothing data, as well as both techniques for comparing time series. The closing Section \ref{sec:results-and-discussion} contains all results, i.e. the answers to the three sub-questions above, and provides a nuanced discussion and a concluding statement. This data analysis paper is followed by our modelling paper \cite{rollier2022b}, which is currently under review.

\section{Data collection}
\label{sec:data-collection}

\subsection{Geographical and temporal demarcations}
\label{subsec:demarcations}

\begin{figure}[h]
    \centering
    \includegraphics[width=.6\linewidth]{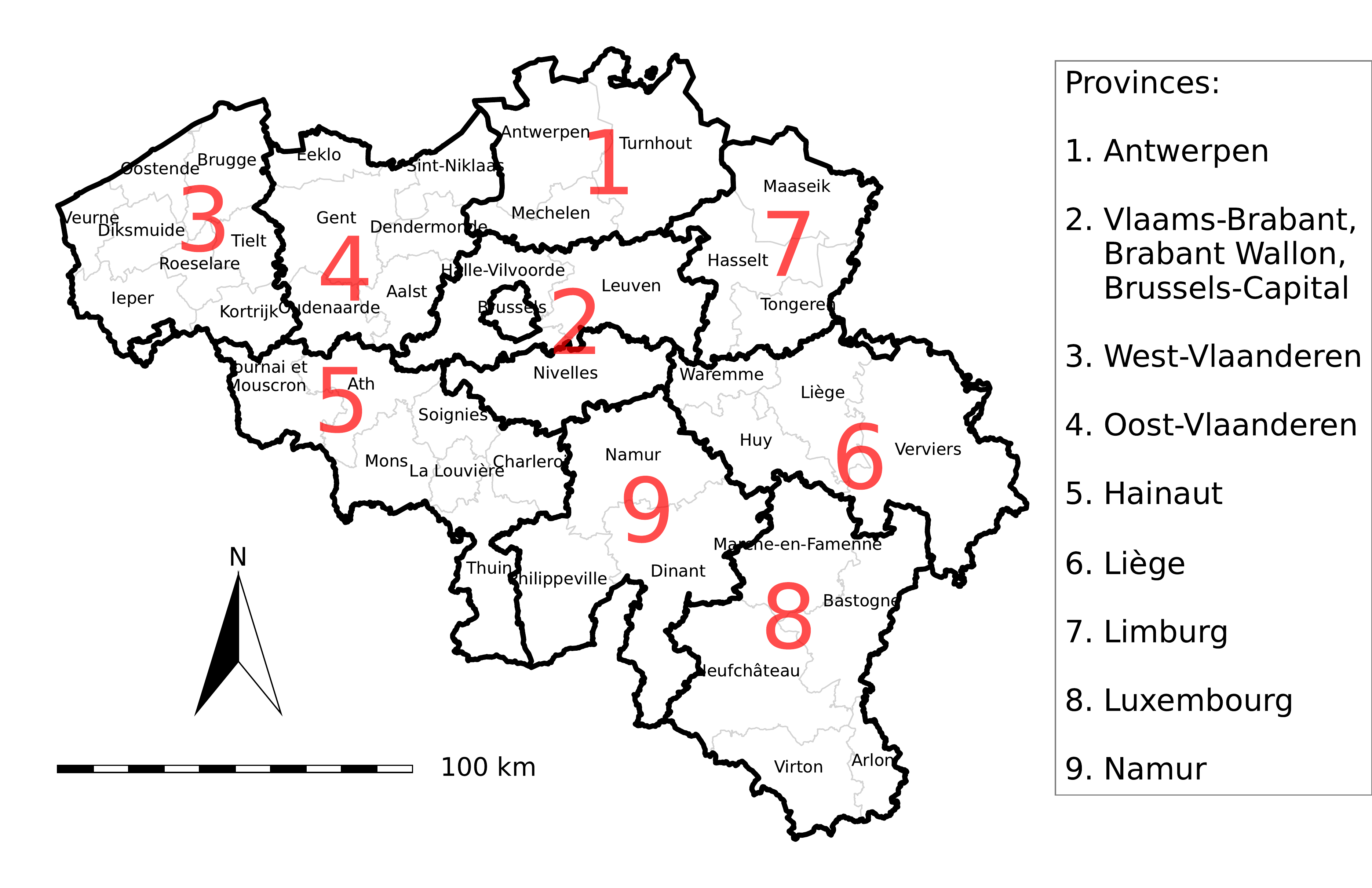}
    \caption{The 43 Belgian arrondissements, accompanied in red by the first digit of their systematic identifier (NIS code), which coincides with Belgian provinces (boxed legend). For an exhaustive list of the geographic and demographic properties, consult Tab.~\ref{SI:tab:arr}.}
    \label{fig:map-of-belgian-arrondissements}
\end{figure}

All analyses in this paper are performed at NUTS 3 level (Nomenclature of Territorial Units for Statistics), corresponding to administrative units with an average population size between \num{150000} and \num{800000}. In Belgium, there are 43 such administrative units, called arrondissements, sometimes referred to as ``districts'' in English (Fig.~\ref{fig:map-of-belgian-arrondissements}, Tab.~\ref{SI:tab:arr}). Three reasons motivate this choice of spatial resolution: 1) we aim for consistency and comparability with other spatial analysis papers within Europe, such as the aforementioned study by Iacus et al.~\cite{Iacus2020}; 2) we avoid troublesome analysis of overly noisy time series associated with smaller geographical units; and 3) we avoid unneeded complication arising from the EU GDPR legislation when working with privacy-sensitive data at higher spatial resolution. When presenting results as heatmaps, we list the 43 arrondissements according to their systematic number (NIS number) where needed, thereby grouping them per province (NUTS 2, see Fig.~\ref{fig:map-of-belgian-arrondissements}).\\

For the purpose of this paper we delineate two 2020 \covid{} waves whose start, middle and end dates apply to time series of all arrondissements. The first wave starts on March 1 (first confirmed \covid{} cases), rises until March 28, and falls until July 1 (start of summer holidays).
The second wave starts on September 1 (reopening of schools and hence a sudden increase in contact dynamics), rises until October 30, and falls until the end of December 2020. Middle dates are those at which the nationwide maximum number of hospitalisations was registered (See the top arrows in Fig.~\ref{fig:timeline_2020}). These limits define three types of wave periods to which we will refer to as the ascending part, the descending part, and the full wave.\\


We focus our analysis on  2020 because Belgium, just as many other countries, was then largely isolated as a consequence of the imposed containment measures, especially during the waves indicated in  Fig.~\ref{fig:timeline_2020}. This unique situation allows to ignore cross-border mobility and hence consider Belgium as a quasi-independent geographic entity.

\subsection{COVID-19-related time series}

We analysed spatio-temporal data on excess deaths, and daily new \covid{}-related hospitalisations, at the arrondissement level. Spatially stratified mortality data are freely available per \textit{week} and per arrondissement \cite{Statbel2020a}. We calculated the excess death time series by simply dividing the local weekly data for 2020 by the average values over the ten previous years, hence presenting it as a comparative fraction. This approach was preferred over directly reported \covid{}-related death data, because attempting to acquire the latter involved inconvenient privacy issues. An additional issue with reported \covid{} mortality is that at least two different conventions are adopted internationally \cite{karanikolos2020}, which makes results dependent on protocol. The fraction of excess deaths, on the other hand, does not depend on any such conventions, and is therefore considered a more objective metric.\\

Data on hospitalisations due to \covid{} are provided by the Belgian public health institute Sciensano. These are known at a \textit{daily} basis, per arrondissement, and (here) normalised per \num{100000} inhabitants. While these detailed data are not freely available, an aggregated format of these datasets is in the public domain \citep{Sciensano2020}; Fig.~\ref{fig:timeline_2020} shows the nationally aggregated hospitalisation time series. Note that we also analysed time series of daily new confirmed \covid{} cases. These are, however, not included here  because they depend  on test capacity. 

\subsection{Mobility data}
\label{subsec:mobility-data}

As a proxy for Belgian mobility, we used positioning data from mobile phones connecting to transmission towers. These data are provided by Proximus, Belgium's largest telecommunication company. Considering that these data account for the movement of 25 to 50\% of the population in any given arrondissement (private communication and Ref. \citep{FOD_economie_proximus_market-share}), we assume the data are representative of the entire population dynamics. The processed time series are expressed as the daily ``staytime'' $P^{gh}(t)$, denoting the total amount of time spent by \textit{all} residents of arrondissement $g$ in arrondissement $h$ during the day corresponding with time $t$. We will only use mobility quantities averaged over any of the four previously defined partial waves, and a prepandemic baseline.

\section{Methods}
\label{sec:methods}


\subsection{Average daily mobility and connectivity index}

To verify that mobility was severely affected during the pandemic (the first objective), we compare the average daily outward mobility during the ascending and descending phases of both the first and second 2020 \covid{} waves with prepandemic mobility, per \num{100000} inhabitants. We compute this as:
\begin{equation}
    (\text{average outward mobility})^g = \frac{1}{\Delta t}\frac{100 000}{N^g}\sum_{t }\sum_{h\neq g}P^{gh}(t),
    \label{eq:avg-mobility-formula}
\end{equation}
where superscript $g$ denotes the considered (home) arrondissement, $h$ the visited arrondissement, $\Delta t$ the number of days in the considered period, and $N^g$ the population of arrondissement $g$. Note that we use \textit{super}scripts for consistency with our other work \cite{Alleman2021, rollier2022b}.\\



We further define a mobility-based connectivity index (CI), inspired by Iacus et al. \cite{Iacus2020}:
\begin{equation}
    \text{CI}^{gh} = \ln\left[\frac{1}{\Delta t_\text{wave}}\sum_{t \in \text{wave}}\left(P^{gh}(t) + P^{hg}(t)\right)\right].
    \label{eq:CI-formula}
\end{equation}
This quantity expresses how well two arrondissements are connected over a certain period of time, irrespective of the  direction of movement, i.e. $\text{CI}^{gh}=\text{CI}^{hg}$ such that it is a property of the (unordered) pair $\{g,h\}$. The staytimes $P^{gh}$ and $P^{hg}$ in Eq.~\eqref{eq:CI-formula} are not normalised in order to arrive at an \textit{absolute} measure of connectivity, rather than at an insight on  how inclided an individual in $g$ is to visit $h$. The natural logarithm is  used for the sake of scaling, but other monotonically increasing functions may suffice as well, considering we will quantify our results with a rank correlation coefficient (see Subsection \ref{subsec:correlating-mobility-and-covid19-time-series}).



\subsection{GAMM fitting and bootstrapping of COVID-19 time series}\label{sec:GAMfit}

The unprocessed \covid{} time series under consideration typically show high variability between days at the arrondissement level. This is partly explained by differences in reporting between week and weekend, and much of the additional variation comes from the inherently stochastic nature of (detecting) infections and hospitalisations. Rather than directly analysing noisy original data, we want to meaningfully compare \textit{trends} between various arrondissements, and simultaneously provide an indication of uncertainty. An elegant way to achieve both goals is to model each time series using a Generalised Additive Mixed Model (GAMM) approach with a log link \citep{Wood2017}.\\

First, \textit{smoothing} is established by modelling the expected count of events $E(Y_t)$ per \num{100000} inhabitants per day $t$ as
\begin{equation}
E(Y_t) = \mbox{exp}(\alpha + \bm{s}(\bm{X})_t\mathbf{\bm{\beta}}) + \epsilon_t.
\label{eq:GAMM-formula}
\end{equation}
In this equation, $\alpha$ represents a general intercept. The vector $\bm{s}(\bm{X})_t$ is the $t$-th row of a matrix containing the projection of time vector $\bm{X}$ on a second-order penalised B-spline basis. We chose the dimension of the spline basis to roughly match the number of weeks in the dataset, which allows for enough detail to incorporate the peaks in number of events, while the penalisation helps avoiding erratic fluctuations \citep{eilers1996}. The dot product in the exponential is with $\bm{\beta}$, a vector of coefficients for the fixed effects.\\

If many people are hospitalised on day $t$, fewer people remain to be hospitalised on day $t+1$, i.e. the data points in the time series are \textit{auto-correlated}. Simultaneously, due to random events cancelling out, we are relatively more certain (in terms of signal-to-noise) of a high number of hospitalisations compared to a low number, i.e. depending on the statistical model, the data points may be \textit{overdispersed}.The GAMM approach outperforms less advanced methods because it allows us to incorporate both autocorrelation and overdispersion. The residuals $\epsilon_t$ are modelled according to a first-order autoregressive process (AR1) to account for auto-correlation between (subsequent) residuals. The estimation itself was performed using a quasi-likelihood method that assumes a linear relation between the mean and the variance to account for the documented overdispersion in \covid{} transmission, as was done by an early Italian study by Scortichini et al. \cite{Scortichini2020}. \\

\begin{figure}[h]
    \centering
    \includegraphics[width=\linewidth]{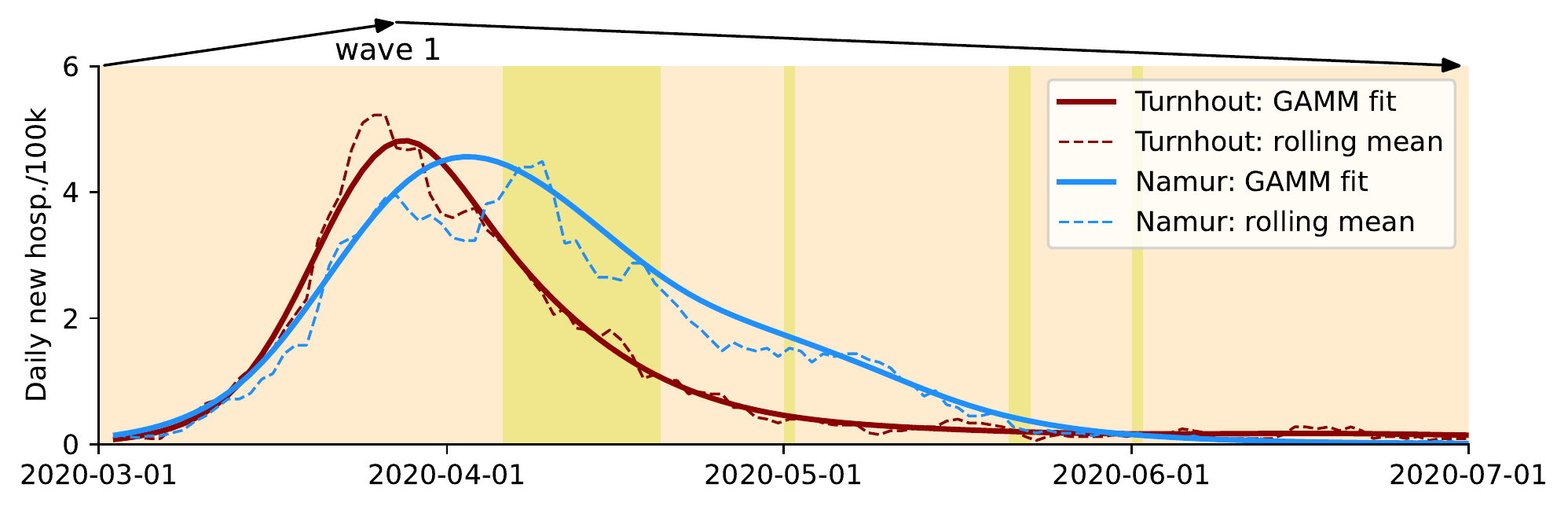}
    \caption{Time series for daily new \covid{} hospitalisations per \num{100000} inhabitants for the arrondissements Turnhout (dark red) and Namur (blue) for the entire first wave. The dashed curves represent the 7-day moving averages. The solid line is a single realisation following the Generalised Additive Mixed Model (GAMM) approach. Per arrondissement, 100 subtly varying GAMM fits are realised, allowing for a systematic quantification of the time series' variability.}
    \label{fig:GAMM-fit-13000-92000_wave1}
\end{figure}

Fig.~\ref{fig:GAMM-fit-13000-92000_wave1} illustrates the result of the GAMM approach for two Belgian arrondissements, Turnhout and Namur, on \covid{} hospitalisation data. This plot also visualises the corresponding rolling average with a one-week window, illustrating the additional ``smoothness'' the GAMM approach provides, while keeping relevant epidemic trends.\\

\textit{Indicating uncertainty} is established by exploiting the stochasticity inherent to the GAMM approach outlined above. For every original time series, we simulate not just 1 but 100 GAMM curves using a parametric bootstrap procedure based on the spline coefficients. In this procedure, we treat the coefficients as originating from a multivariate normal distribution, with the estimated value as mean and the estimated variance-covariance matrix of these coefficients as the variance structure. For every simulated time series a vector with coefficient values $\bm{\beta}'$ is randomly drawn from this distribution, and multiplied with the matrix $\bm{s}(\bm{X}_t)$. Basically, original time series with a higher signal to noise ratio due to a limited number of daily counts will naturally produce more variability over the 100 GAMM fits, and hence more deviation over the resulting analysis values. We again refer to \cite{Wood2017} for details, and to Fig.~\ref{fig:GAMM-fits-examples} for an example of the resulting ``spectrum'' of fits.



\subsection{Comparison between time series}

\begin{figure}
    \centering
    \includegraphics[width=0.45\linewidth]{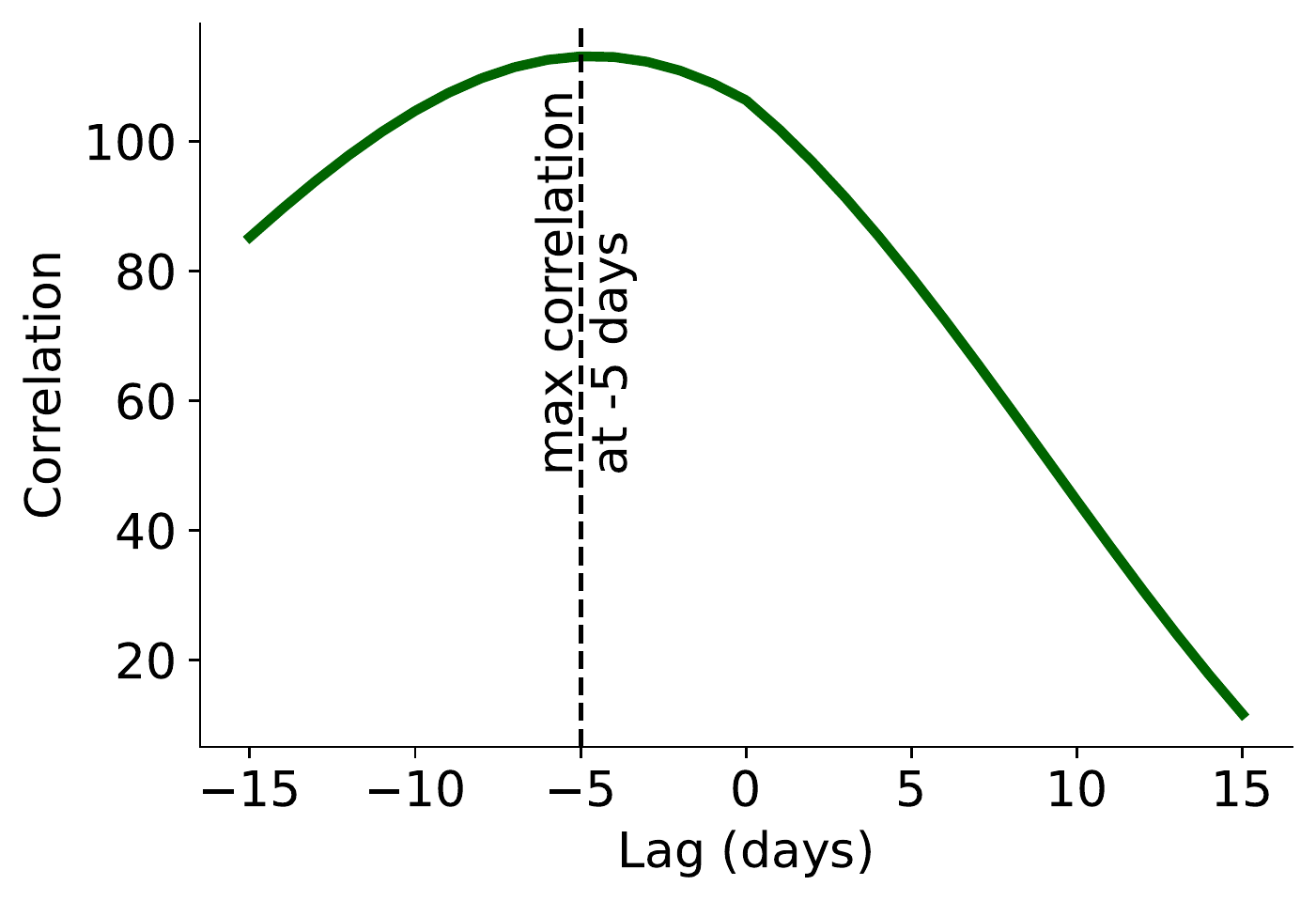}
    \includegraphics[width=0.43\linewidth]{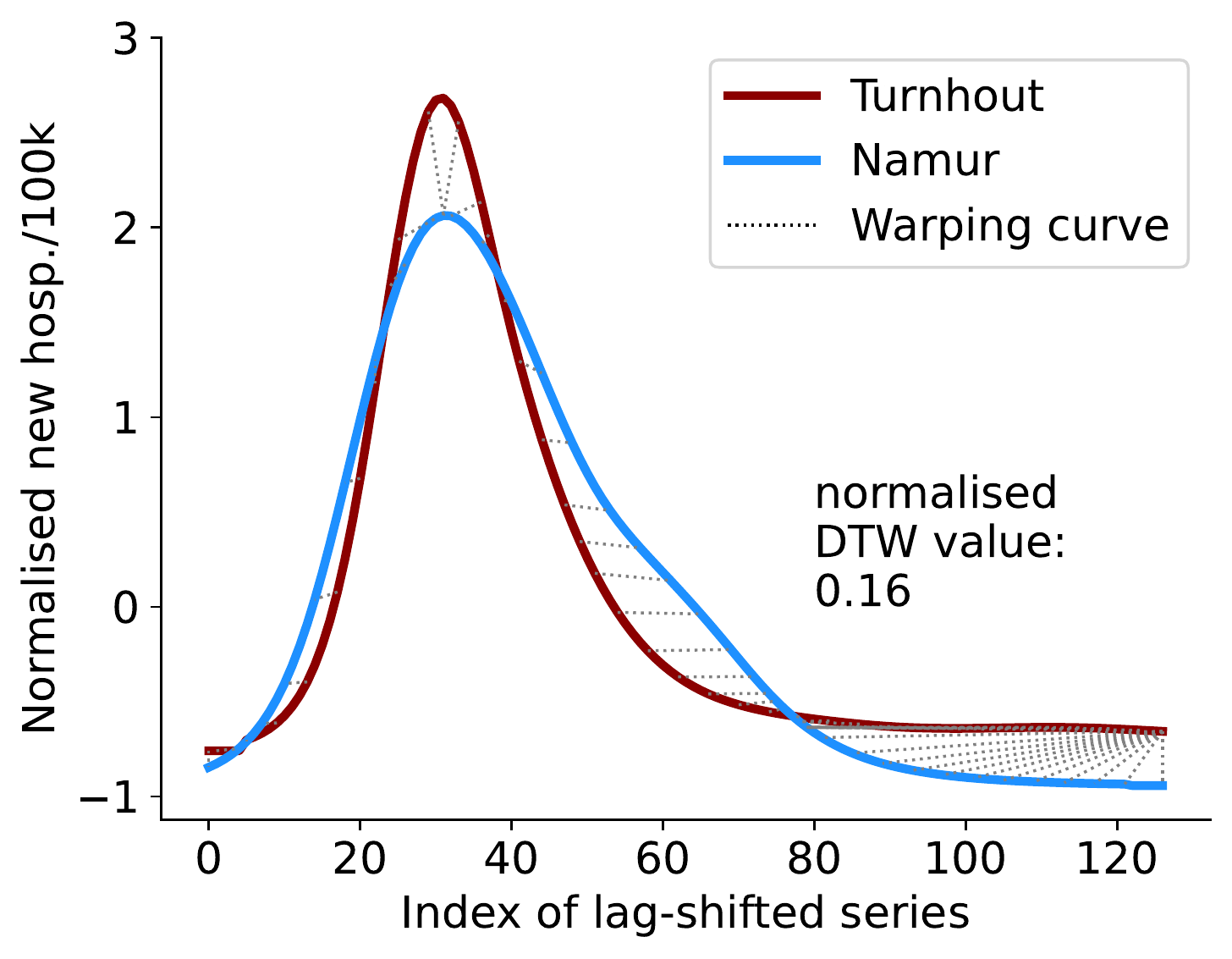}
    \caption{\textit{Left}: Time-lagged cross-correlation values for time lags ranging between $-15$ and $15$ days between the GAMM-fitted hospitalisation time series for Turnhout and Namur shown in Fig.~\ref{fig:GAMM-fit-13000-92000_wave1}. 
    \textit{Right}: illustration of the alignment enforced by the dynamic time warping between these time series, resulting in a normalised DTW value of 0.16.}
    \label{fig:TLCC-and-DTW_example-13000-92000}
\end{figure}

Local \covid{} time series generally resemble the epidemic behaviour at the national level, but we expect minor differences between arrondissements in amplitude and timing to be indicative of the geographical direction and intensity of the viral spread. More concretely, we anticipate that strongly connected arrondissements have near-synchronous time series with a highly analogous shape. We systematically compare hospitalisation time series for all unique arrondissement couples, for both the first and the second 2020 \covid{} wave. We assess \textit{morphological similarities} between smoothed time series using dynamic time warping (DTW). \textit{Difference in timing} (lag) is computed using time-lagged cross-correlation (TLCC). Both methods, which are briefly explained below, first impose standardisation to zero mean and unit standard deviation (``standard-score normalisation''). The significance of the resulting values is gauged by looking at the average values and standard deviations over all 100 GAMM fits.\\

The TLCC quantifies how synchronised two time series $s^g(t)$ and $s^h(t)$ are by computing the correlation $C$ between them for different time shifts. The highest correlation indicates the time lag for which both signals match up best \citep{shen2015analysis, paploski2016time}. That is to say, we look for the (negative or positive integer) value $k$ for which
\begin{equation}
    C_{s^g s^h}(k) = \sum_{t \in \text{wave}} s^g(t+k)s^h(t)
    \label{eq:TLCC}
\end{equation}
is maximal. Here $s^g$ and $s^g$ are the standard-score normalised \covid{} time series for arrondissements $g$ and $h$, which are here defined to vanish outside the considered time period. We are not interested in the value of $C_{s^g s^h}(k)$, which has little physical importance, but only in the horizontal position of the maximum, symbolised simply by
\begin{align*}
\text{TLCC}^{gh} = \text{arg max}_k (C_{s^g s^h}(k)).
\end{align*}
For instance, Fig.~\ref{fig:GAMM-fit-13000-92000_wave1} suggests visually that for these two GAMM realisations, the peak of the daily new hospitalisations in Turnhout occurs earlier than in Namur. This is reflected by the maximum TLCC occurring at a lag of -5 days (Fig.~\ref{fig:TLCC-and-DTW_example-13000-92000}, left), which allows us to conclude that Turnhout was about 5 days ahead of Namur for what concerns the time series for confirmed cases. Note however that the 99 other GAMM fits may generate different lag values, so the resulting value distribution's standard deviation provides an indication for uncertainty. Additionally, note that this result, regardless of whether it is significantly non-zero, does not imply any type of causation.\\

DTW gauges morphological similarity by minimising the Euclidean distance between two standard-score normalised time series $s^g(t)$ and $s^g(t)$ of duration $\Delta t$, through the deformation (remapping) of the time indices, after synchronisation based on TLCC lags \citep{Kongming1997}. The latter precondition is important in order to ascertain that the DTW value (often called DTW distance) does not \textit{intrinsically} correlate with the TLCC value, hence safeguarding that both values communicate independent information. As the DTW distance scales with the number of analysed data points, we must normalise over the number of days $\Delta t$ of the considered period for better comparison between distinct waves. Mathematically, we find a warping curve $\phi(k)$ ($k \in \{1, ..., T\}$)
\begin{align*}
    \phi(k) = \left(\phi_{s^g}(k), \phi_{s^h}(k)\right), \text{ with }
    \phi_{s^g}(k), \phi_{s^h}(k) \in \{1, ..., \Delta t\}
\end{align*}
which allows to minimise
\begin{equation}
    \text{DTW}^{gh} = d_\phi(s^g,s^g) = \frac{1}{\Delta t}\sum_{k=1}^T \lvert\phi_{s^g}(k) - \phi_{s^h}(k)\rvert m_\phi(k) / M_\phi,
    \label{eq:DTW}
\end{equation}
where $m_\phi(k)$ is a per-step weighting coefficient and $M_\phi$ is the corresponding normalisation
constant. We refer to Giorgino (2009) \cite{giorgino2009computing} for details on the algorithm. Minimising the DTW value can be thought of as stretching or compressing the second series $s^h$ with the aim of having it resemble as much as possible the first (reference) series $s^g$. We are again not interested in the numerical value of this quantity per se, but rather in the comparison between distributions of such values for the 100 GAMM realisations of different arrondissement pairs, where lower DTW values indicate higher similarity. An illustration of the time axis deformation is shown in the right panel of Fig.~\ref{fig:TLCC-and-DTW_example-13000-92000}.

\subsection{Correlating mobility and COVID-19 time series}
\label{subsec:correlating-mobility-and-covid19-time-series}

Third, we investigate the relation between mobility and \sars{} spread, by first looking at \textit{particular} early-outbreak arrondissements and associated excess deaths, and then by \textit{generally} investigating the correlation between CIs on the one hand, and DTW values and TLCC lags on the other.\\

We anticipate that Belgium follows the trends observed in Ref. \cite{Iacus2020}: a strong connection to an early-outbreak arrondissement on average foretells an increased excess mortality some weeks later. We aim to verify this by first identifying three early-outbreak arrondissements. For each of these arrondissements, we plot the associated first-wave ascending-phase CIs to all other arrondissements, and the average percentage of weekly excess deaths in these arrondissements, for every first-wave week since the start of the outbreak. The resulting time-dependent correlation between CIs and excess deaths (ED) for all $G-1=42$ arrondissements connected to arrondissement $g$ is expressed as the Spearman's rank correlation coefficient $\rho$ \cite{spearman1987proof}. This non-parametric approach enables the detection of \textit{any} monotonic relationship between both variables, which is desirable due to our agnosticism regarding the precise nature of the relation. Values closer to 1 (resp. -1) indicate stronger correlation (resp. anticorrelation).\\


Next, with \textit{all} unique 903 CI values from Eq. \eqref{eq:CI-formula} on the one hand, and \textit{all} 903 TLCC lags and DTW values from Eqs. \eqref{eq:TLCC} and \eqref{eq:DTW} on the other, we construct a number of scatter plots in which we quantify correlations between data pairs $(\text{CI}^{gh}, \vert\text{TLCC}^{gh}\vert)$ and $(\text{CI}^{gh}, \text{DTW}^{gh})$. Note that we consider the absolute value for TLCC lags, because we are only interested in the \textit{magnitude} of the time shift between the involved time series. We generally anticipate negative correlations: strongly connected arrondissements (high CI values) are expected to demonstrate small epidemiological time lags (small TLCC lags) and morphologically similar time series (small DTW values). We again calculate Spearman's rank correlation coefficients. Values are calculated for the ascending, descending, and full period of both 2020 Belgian \covid{} hospitalisation waves based  on the respective GAMM fits (100 per time series).


\section{Results and discussion}
\label{sec:results-and-discussion}

\subsection{Mobility changes during the 2020 COVID-19 waves}
\label{sec:mobility_changes}

\begin{figure}[h]
    \centering
    \includegraphics[width=.60\textwidth]{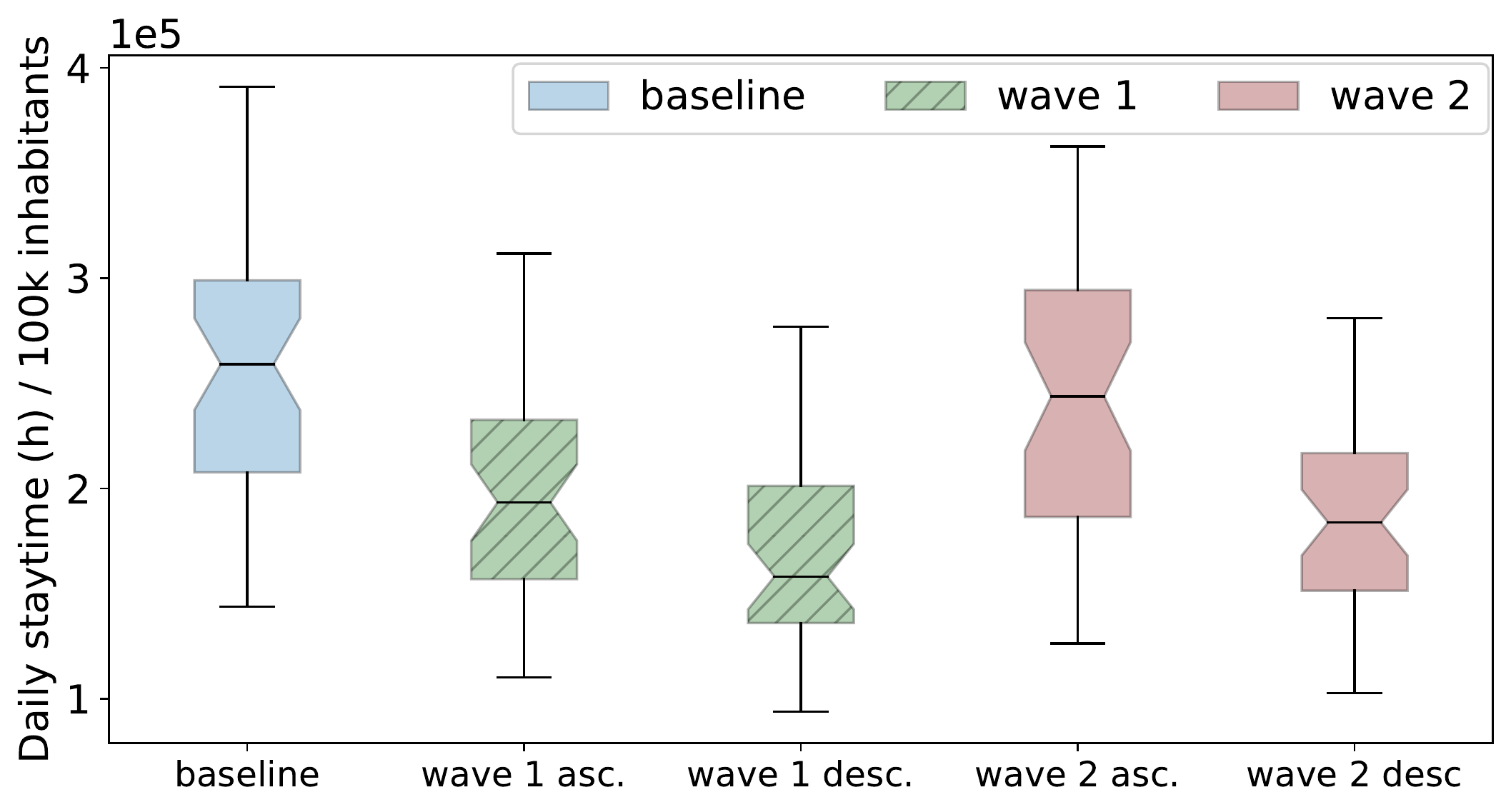}
    \caption{Boxplots of daily mobility of all 43 arrondissements, calculated from Eq. \eqref{eq:avg-mobility-formula}} as the total number of hours spent outside the home arrondissement per \num{100000} inhabitants, averaged over the considered period. The baseline mobility is calculated over the period 20-28 February 2020.
    \label{fig:boxplots}
\end{figure}

\begin{figure}[h]
    \centering
    \includegraphics[width=.80\textwidth]{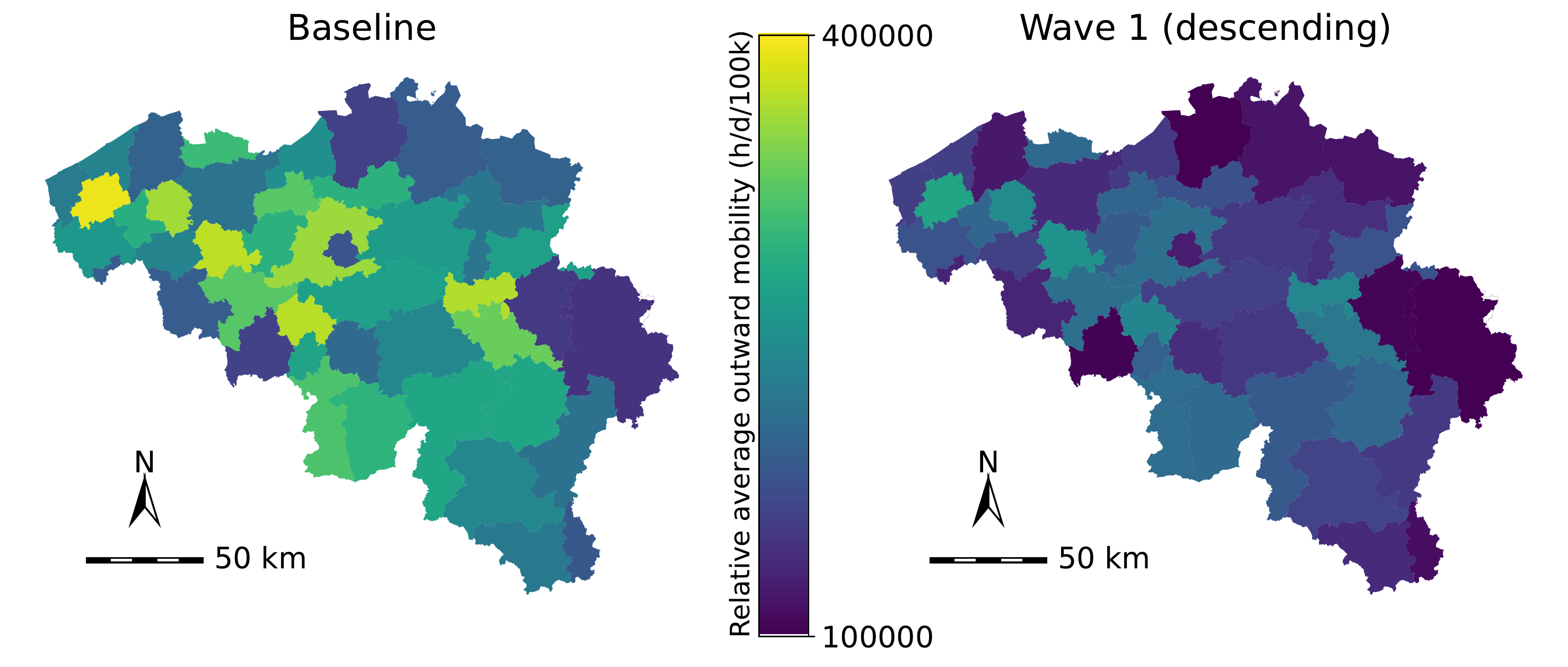}
   \caption{Maps of Belgian arrondissements, coloured according to the average outward mobility during the prepandemic baseline period (left), and during the descending part of the first wave (right).}
   \label{fig:thematic-maps}
\end{figure}

Confirming intuition, the average mobility outside the home arrondissement dropped substantially during the ascending and descending phases of both \covid{} waves in comparison to the baseline mobility. This is shown clearly in the boxplots for the five chronologically ordered periods in Fig.~\ref{fig:boxplots}, and in the geograpically detailed thematic maps in Fig.~\ref{fig:thematic-maps}. In particular, we observe that the change in outward mobility is less pronounced for the ascending parts compared to the ascending parts of both waves. This makes sense, considering that the lockdown period during both waves mainly coincides with their descending phases (see dashed double arrows in Fig.~\ref{fig:timeline_2020}).
Additionally we observe that overall average outward mobility was higher for the second wave compared to the first.
\\

\begin{figure}[h]
    \centering
    \includegraphics[height=.35\textwidth]{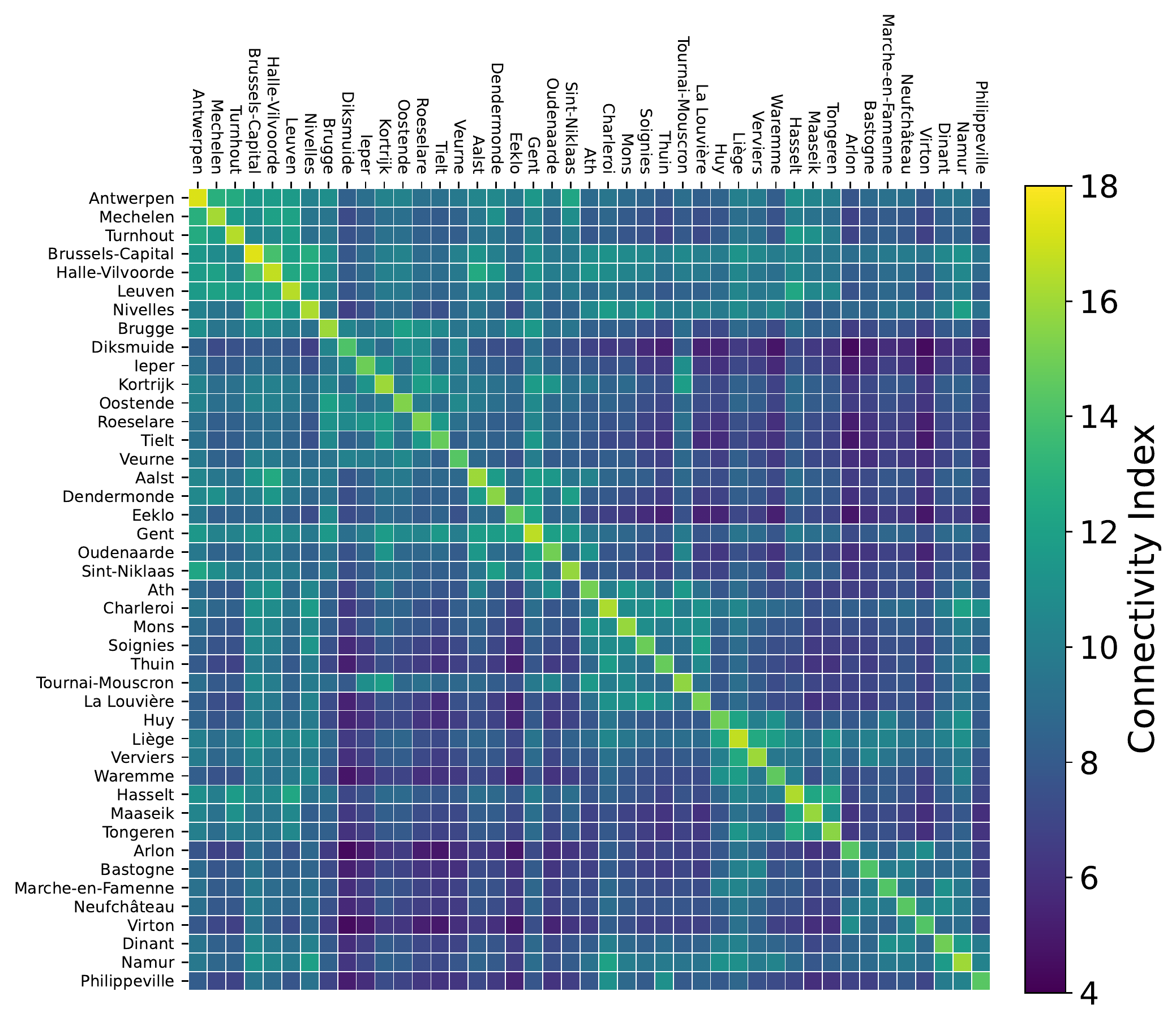}
    \includegraphics[height=.35\textwidth]{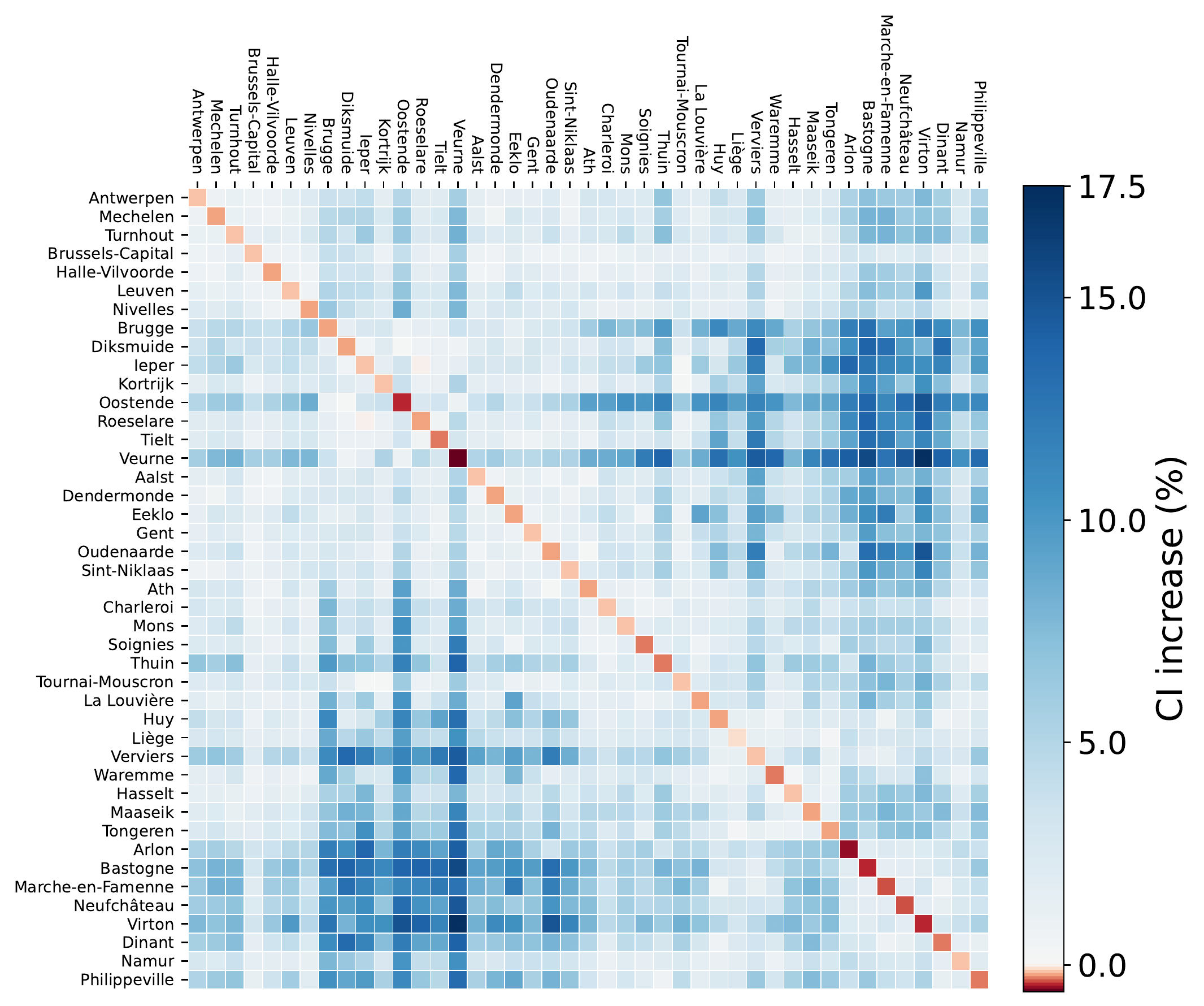}
    \caption{\textit{Left}: symmetric heatmap showing the first-wave connectivity indices (CIs), defined in Eq. \eqref{eq:CI-formula}, ordered in geographical clusters for increasing NIS identifier (see Tab.~\ref{SI:tab:arr}). The highest values at the diagonal show that most time is consistently spent in the home arrondissement. \textit{Right}: The second heatmap shows the difference in CI between the first and second 2020 \covid{} wave in terms of percentage. Positive percentages at the diagonal indicate a rise in CI during the second wave. The full CI heatmap for the second wave is found in Fig.~\ref{SI:fig:heatmap_wave2}.} 
    \label{fig:heatmaps_CI}
\end{figure}

The connectivity indices defined in Eq. \eqref{eq:CI-formula} demonstrate mobility changes in a more fine-grained fashion: (symmetric) heatmaps containing CIs are shown and compared for the two waves in Fig.~\ref{fig:heatmaps_CI}.
From this heatmap we infer that all but one arrondissements containing a province capital are relatively well connected to any other arrondissement according the CI metric, which is of course mainly due to the fact that such arrondissements typically have more inhabitants. Arlon is the only exception to this rule, which is expected because of its location, its small population of some \num{63000}, and the fact that a considerable number of its inhabitants normally commute abroad to the Grand Duchy of Luxembourg. This is in agreement with former studies on the general metropolitan connectivity in Belgium \citep{vanmeeteren2016} and in line with the \covid{}-related comprehensive work by Islam et al. \cite{Islam2021}.
Comparing the first and second wave, the change in CI was relatively consistent between most arrondissement pairs: a small 5 to 10\% increase during the second wave. Notable exceptions are connections with arrondissements in the province of West-Vlaanderen to the arrondissements Arlon, Bastogne, Virton, Veurne and Philippeville (over 15\%). Despite highlighting the connectivity changes, the relevance of the actual CI values is limited due to the CI's pragmatic definition. Still, they  become meaningful in comparison to the corresponding DTW and TLCC values, to which we turn next.

\subsection{Spatio-temporal dynamics of the 2020 COVID-19 waves}

\begin{figure}[h]
    \centering
    \includegraphics[width=0.9\textwidth]{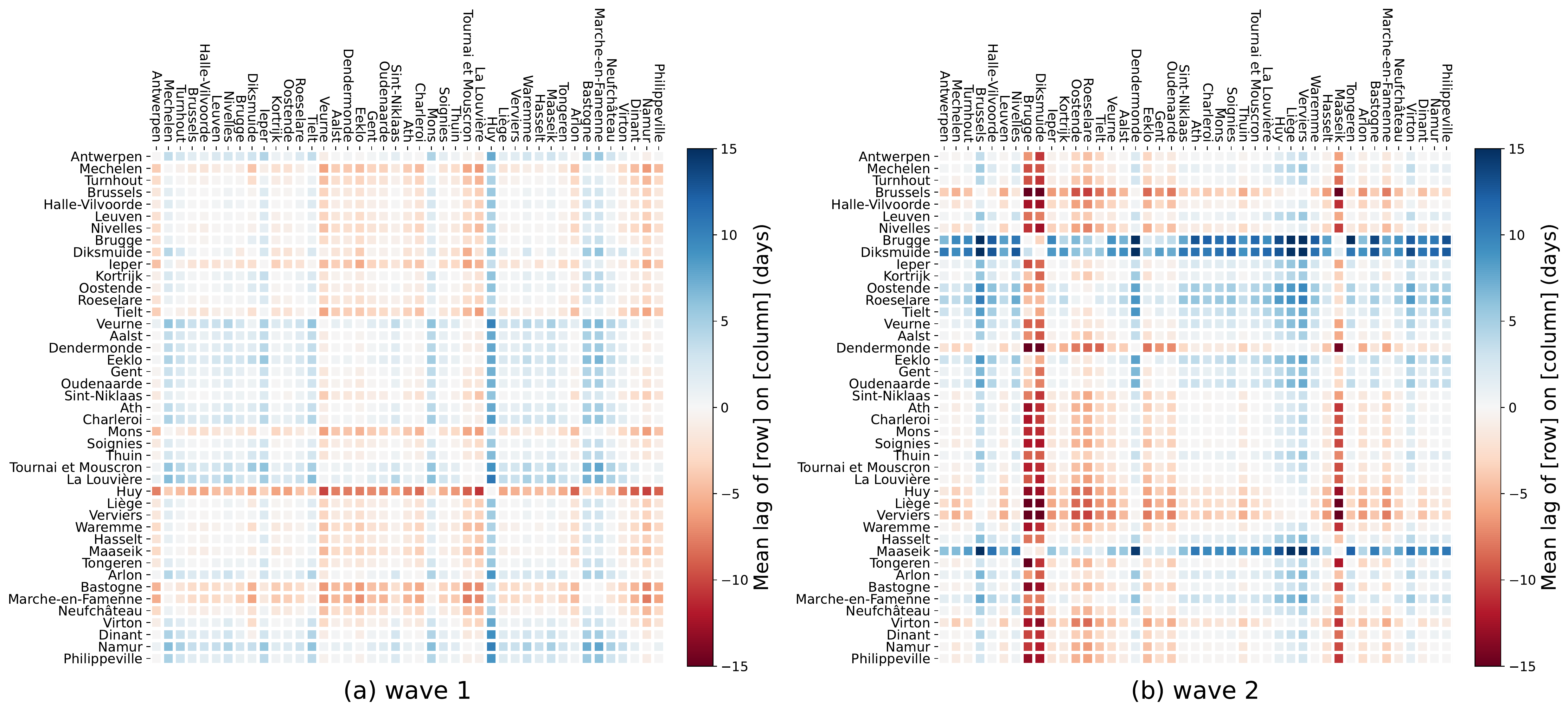}\\
    \includegraphics[width=0.9\linewidth]{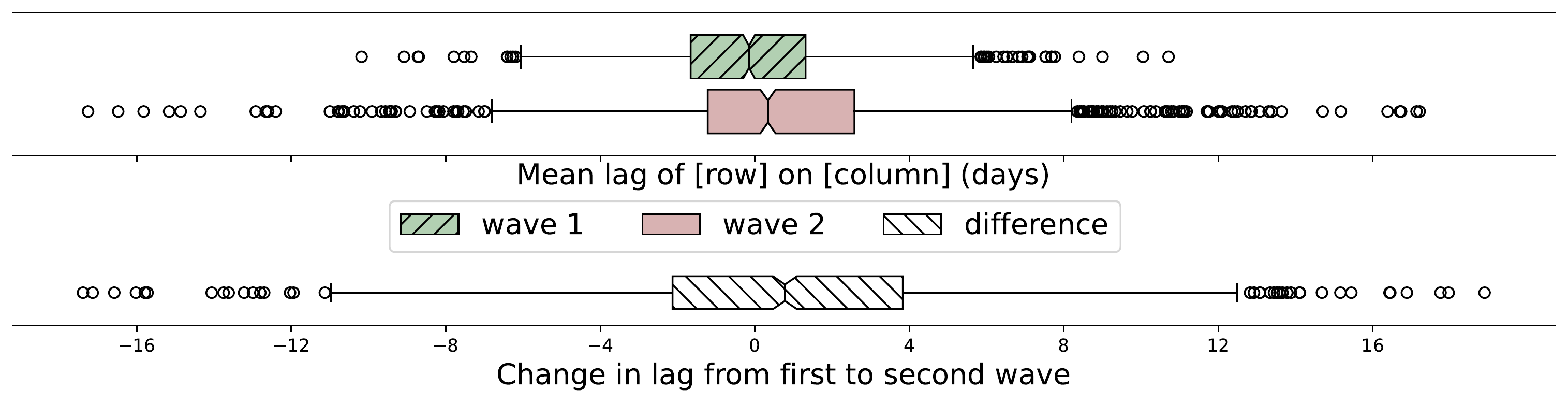}
    \caption{\textit{Top:} Average time-lagged cross-correlation lag computed between each pair of Belgian arrondissements from 100 GAMM fits on the hospitalisation time series, for first (a) and second (b) \covid{} waves. The colour scale indicates how many days (on average) the arrondissement indicated by the row header lags on the one indicated by the column header. \textit{Bottom:} Identical TLCC information of all 903 unique arrondissement pairs, plotted in boxplots. There is a clear change between the first and the second wave, as shown by the bottom boxplot, with a median value of differences that is significantly greater than zero (Wilcoxon test $p$ value $\sim10^{-6}$)}. Additional mean TLCC lag values as well as plots with standard deviations are found in Supplementary Material.
    \label{fig:TLCC_hosp}
\end{figure}

The heatmaps in Fig.~\ref{fig:TLCC_hosp} show the average time lag (in days) over all bootstrapped hospitalisation time series of both the first and second full \covid{} wave, for each pair of arrondissements. The corresponding standard deviation can be found in Fig.~\ref{fig:TLCC_hosp_std}. From both figures, we  infer for instance that the \covid{} hospitalisation wave in Brussels-Capital during the first 2020 wave was about $1.6\pm0.7$ days (significantly) ahead of the one in the Antwerpen arrondissement, and $0.5 \pm 9.8$ days (insignificantly) ahead of Diksmuide arrondissement.\\


According to Sciensano \citep{Sciensano2020}, the regions in the southwest of province Limburg and the so-called Borinage region are defined as the initial clusters in Belgium. These regions correspond with the arrondissements Tongeren and Hasselt (Limburg), and Mons (Borinage). Upon inspection of Fig.~\ref{fig:TLCC_hosp}, this antecedence is most clearly visible for Mons, which has a negative lag on virtually all other arrondissements, and is a maximum of $6.3\pm1.3$ days ahead compared to Namur. For Tongeren and Hasselt this is less convincingly so, arguably because the TLCC gauges the lag of the entire wave, while Sciensano identified initial clusters based on local index patients. We also notice that small arrondissements appear to have larger TLCC lags, which is presumably an artefact of the relatively large noise on these time series and the resulting GAMM fits.
This is particularly visible for the arrondissements Diksmuide and Ieper during the first wave and translates to a very high TLCC lag standard deviation (see Figs.~\ref{fig:GAMM-fits-examples} and \ref{fig:TLCC_hosp_std}), and indicates that performing this analysis on even smaller geographical units would probably lead to meaningless results.

Interestingly, a clear distinction can be made between both \covid{} waves when it comes to TLCC lags. Focusing on the larger geographical units, Brussels and Liège are now clearly ahead, and the province of West-Vlaanderen as a whole appears to lag behind. Generally, the lags observed during the second wave are overall larger than the ones observed during the first wave. This is seen in the boxplots in the bottom panel of Fig.~\ref{fig:TLCC_hosp}, and quantified by a one-sided Wilcoxon signed-rank test \cite{Wilcoxon1945}, determining that the median of the differences is greater than zero with high confidence ($p\sim10^{-6}$). Additionally, the second-wave heatmap shows higher variability; this can be understood as the result of decreased mobility when compared to the \textit{ascending} phase of the first wave (Fig.~\ref{fig:boxplots}) -- despite  national homogenising during the summer and overall increased mobility during the \textit{entire} wave. This suggests that a spatial analysis is especially feasible during times of low mobility, and already hints at a link between viral spread and population dynamics.\\

\begin{figure}[h]
    \centering
    \includegraphics[width=0.9\textwidth]{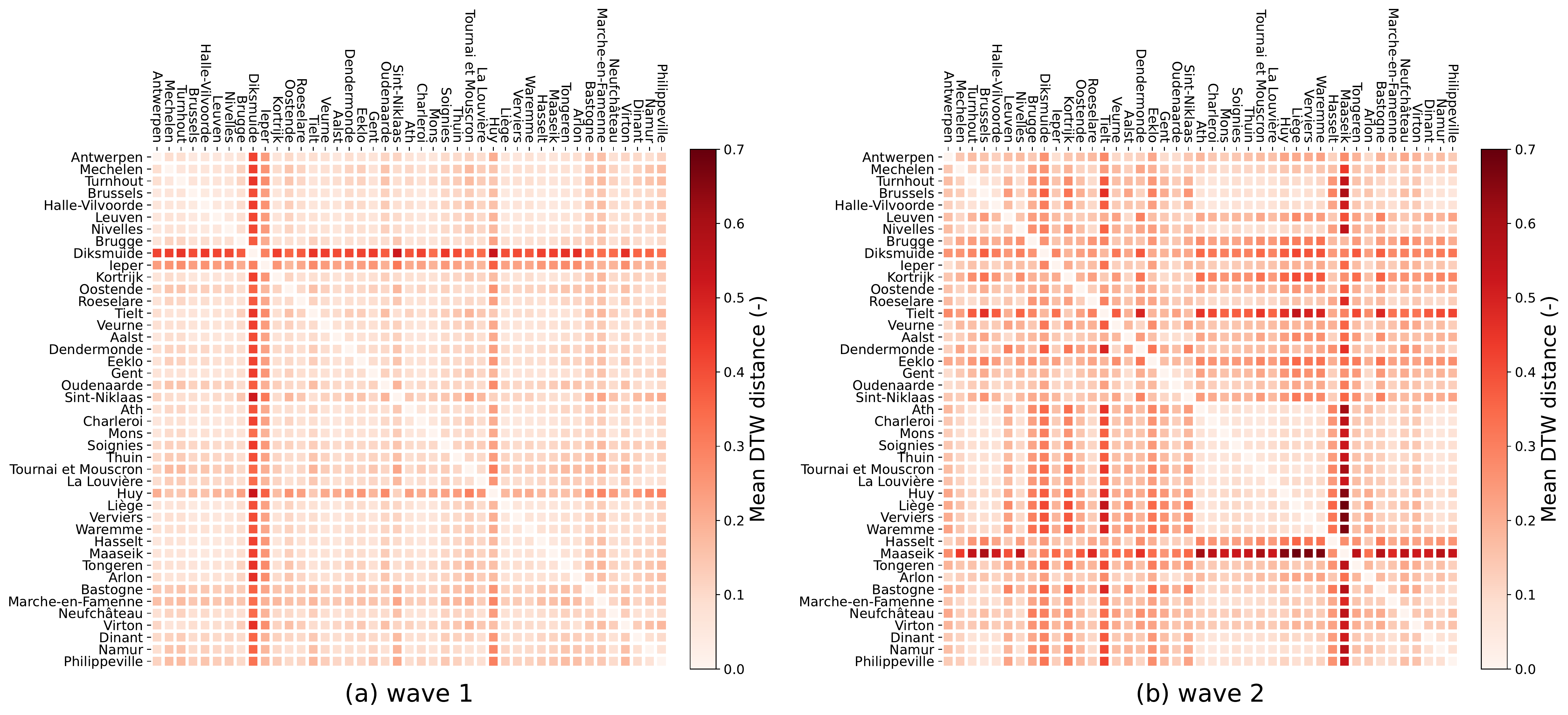}\\
    \includegraphics[width=0.9\textwidth]{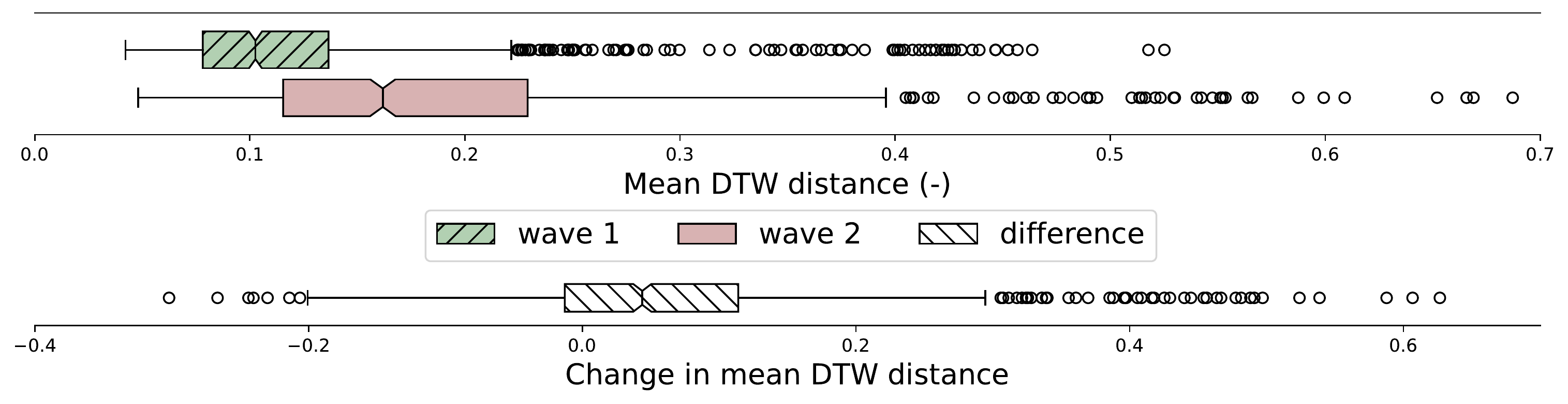}
    \caption{\textit{Top:} Average normalised dynamic time warping distances computed between each unique pair of Belgian arrondissements from 100 GAMM fits on the hospitalisation time series, for first (a) and second (b) \covid{} waves. \textit{Bottom:} identical DTW information, formatted as in Fig.~\ref{fig:TLCC_hosp}. The median value of the differences between wave 1 and 2 is significantly greater than 0 (Wilcoxon test $p$ value $\sim 10^{-73}$).}
    \label{fig:DTW_hosp}
\end{figure}

Similarly, Figs.~\ref{fig:DTW_hosp} (resp. \ref{fig:DTW_hosp_std}) provide a comparison between arrondissements for both 2020 \covid{} waves in terms of the average DTW distance (resp.~standard deviations) over all GAMM realisations, between the corresponding hospitalisation time series.

For the first wave, the largest DTW distances are observed for low-population arrondissements.
This can be understood as confirming our conjecture; these are of course also the least connected arrondissements (Fig.~\ref{fig:heatmaps_CI}). However we must cautiously keep in mind that the GAMM procedure allows for high standard deviations when applied to noisy time series. For the second wave,
generally, highly-populated and geographically close regions have time series that are similar in shape, as is seen by rectangular ``clearings'' around the diagonal in the heatmap. As was observed for TLCC lags, generally the DTW distances have grown for the second \covid{} wave, and more variation between arrondissements is observed.


\subsection{Connectivity and the initial spread of COVID-19 in Belgium}

\begin{figure}
    \centering
    \includegraphics[width=.9\linewidth]{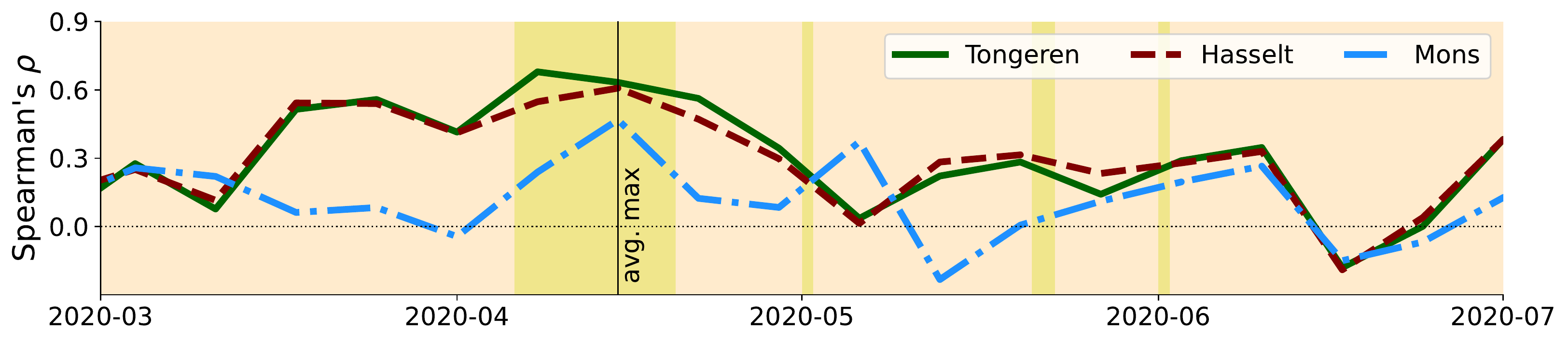}\\
    \includegraphics[width=.9\linewidth]{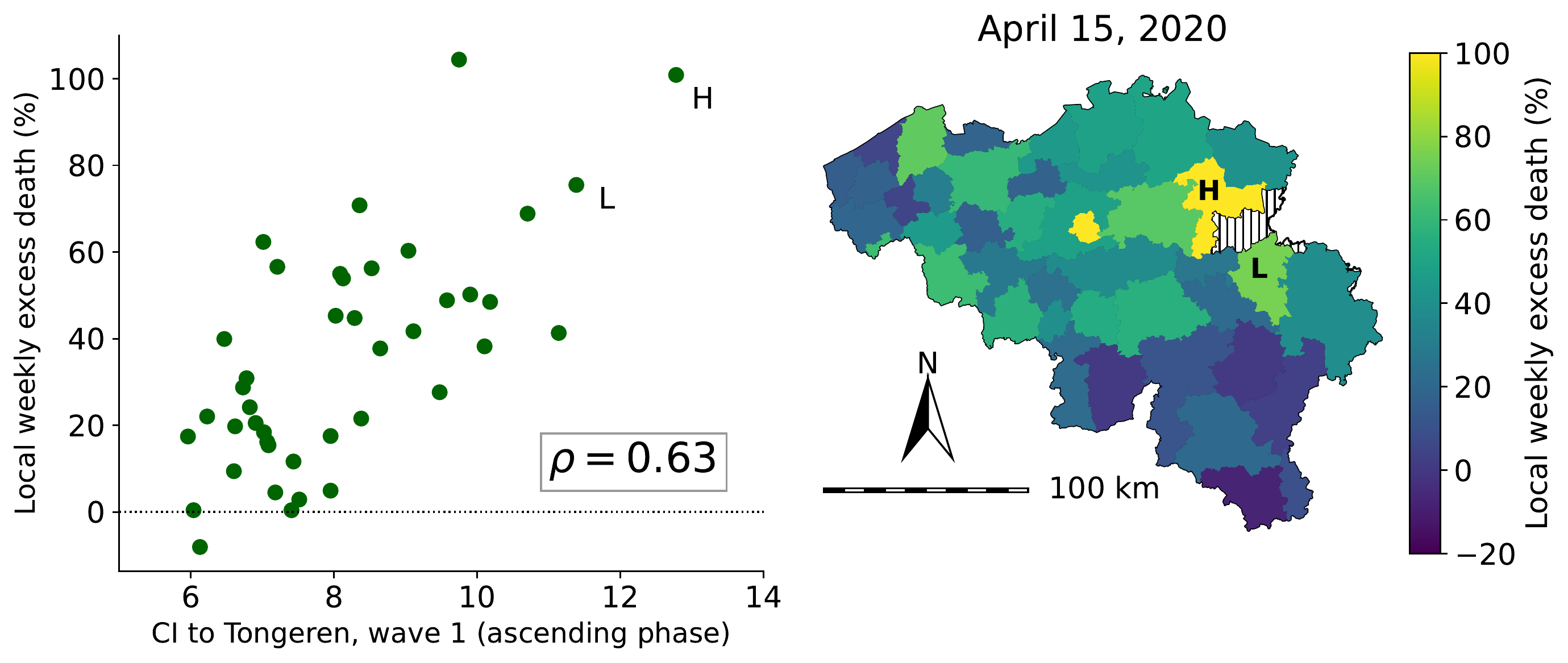}
    \caption{\textit{Top:} Time-dependent correlation between the connectivity indices to early-outbreak arrondissements Tongeren (green, solid), Hasselt (maroon, dashed), and Mons (blue, dot-dashed) on the one hand, and weekly-averaged local excess mortality on the other. The correlation for a monotonic relation is expressed by Spearman's rank correlation coefficient $\rho$}. \textit{Bottom:} Scatter plot and geographical representation of correlation for Tongeren (hatched) during the week of April 15, 2020 (vertical line in upper plot). High-CI arrondissements Hasselt and Liège are indicated with their initials.
    \label{fig:CI-vs-ED}
\end{figure}

We consider Tongeren, Hasselt and Mons as early-outbreak arrondissements, and plot the time-dependent correlation quantities in the top panel of Fig.~\ref{fig:CI-vs-ED}. Clearly, the correlation coefficients for CI-versus-excess-death plots goes up for arrondissements connected to Tongeren and to Hasselt (green and maroon curves) approximately two weeks after the defined start of the first \covid{} wave. It remains quite constant, with a peak Spearman's $\rho$ of resp. 0.63 (Tongeren) and 0.59 (Hasselt) some six weeks into the first wave. This indicates that arrondissements in our dataset that are well connected to these early-outbreak arrondissements, experience a higher excess mortality, which is according to expectation. This is best illustrated for Tongeren on April 15th in the map of Belgium in the bottom right panel of Fig.~\ref{fig:CI-vs-ED} -- see the right panel of Fig.~\ref{SI:fig:heatmap_wave2} for the CIs to Tongeren. The correlation coefficients time series for Mons remain noisy (blue curve). Despite being identified as an early-outbreak arrondissement by Sciensano \cite{Sciensano2020}, we do not see antecedent viral behaviour; nor in the TLCC lag analysis (Fig.~\ref{fig:TLCC_hosp}), nor in this CI-versus-excess-mortality analysis. In any case, from May 2020 onward, the correlation coefficients become negligible and often even negative, demonstrating that connectivity index to the initial hubs is no longer correlated to local excess death: the virus spread has become nationally homogeneous.\\

The influence of connectivity to the French Haut-Rhin department on the initial spread disappeared 14 days after the first lockdown measures \citep{Iacus2020}, which corresponds to the period between first symptoms and death \citep{Lauer2020}. The additional four weeks of delay in response of the virus spread to mobility changes in Belgium can be explained as follows: in contrast to the case of the Haut-Rhin department in France \cite{Iacus2020}, no single big spreading event was documented in one of the initial clusters in Belgium, such that the effect on the percentage of excess deaths was more gradual. Furthermore, Belgium is a very connected country, and hence the difference between most and least connected arrondissements is smaller than in France.

\subsection{Connectivity and COVID-19 dynamics}

\begin{figure}
    \centering
    \includegraphics[width=.9\linewidth]{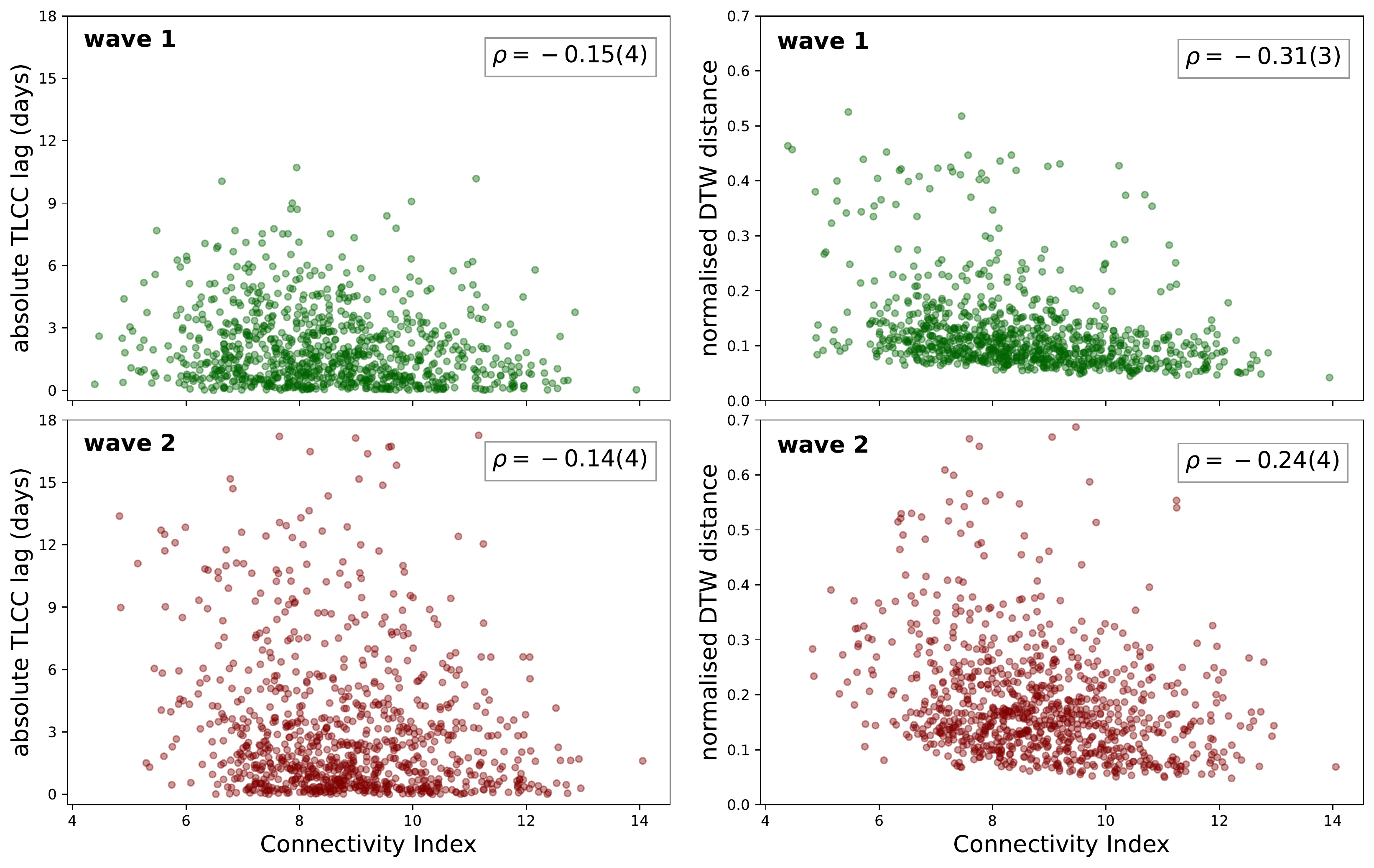}
    \caption{Scatter plots showing all 903 unique arrondissement pairs during the first (top, green) and second (bottom, maroon) wave. We plot the pair's CI values against the normalised TLCC absolute lags (left) and DTW distances (right) from \covid{} hospitalisation time series, averaged over all 100 GAMM fits. For each scatter plot, the average Spearman's rank correlation coefficient $\rho$ is provided with its standard deviation.}
    \label{fig:CI-vs-DTW-TLCC}
\end{figure}

The general results showing correlations between CIs on the one hand, and DTW values or TLCC lags on the other, for first and second \covid{} hospitalisation waves, are shown in Fig.~\ref{fig:CI-vs-DTW-TLCC}. The highest Spearman's rank correlation coefficients are retrieved for the DTW values, in particular for the first wave. The coefficients always point towards a negative correlation, albeit rather weak. The negative correlation coefficients endorse the conjecture that mobility is related to the spatial spread of \covid{}: strongly connected arrondissements will in general exhibit smaller delays (higher TLCC lags) and higher similarity in epidemiological progression (lower DTW values) when compared to poorly connected arrondissements. The scatter plots as well as the values of the correlation coefficients however suggest that the CI cannot be the only predictor for the velocity and morphology of \covid{} spread. In order to complement the results shown in Fig.~\ref{fig:boxplots}, the correlation analyses have been accomplished for the ascending and descending parts of the waves as well (Tab.~\ref{tab:additional_correlation_results}), resulting in lower correlation coefficients. This demonstrates that the analysis of the entire wave is needed to properly assess relations between morphology, synchronicity, and connectivity.\\

The results are in line with those from the study by Habib et al. \cite{Habib2021}, where a (non-linear) spatial linkage between \covid{} and mobility is observed for Belgium as a whole. The correlations at a more fine-grained geographical level considered in this paper, are however weaker. This is in part due to noise of the involved data, but there are several other mobility-related factors that must have also played a role, and that could not be included in our analysis since relevant data are not available at the appropriate spatio-temporal resolution. Of course, the more fine-grained the geographical level, the more the latter are needed for a comprehensive analysis. These factors include the reason behind a mobility event (work, leisure, education, \ldots), the number of stops until the final destination of one mobility event, and so on. Furthermore, by working at NUTS 3 level, we ignored the role of short-distance mobility (mobility within an arrondissement) due to, for instance, local shopping, leisure and educational activities. Taking those short-distance movements into account, would probably allow us to identify a stronger relationship between mobility and \covid{} spread, as also Van De Vijver et al. (personal communication) pointed out that the spatial autocorrelation between \covid{} incidence dropped beyond 15 kilometres for the first wave -- while mostly exceeding 50 kilometres during the second. This again indicates that the assumption of homogeneity within the arrondissements seems to be more valid at the start of the pandemic. Still, an even more fine-grained analysis would require the use of \covid{}-related time series at municipal level, which are generally very noisy due to the relatively low number of hospitalisations per 
municipality, and hence complicating further statistical analysis without gaining much insight.

\section{Conclusion}

Results in this paper were presented in three stages that built up to the main conclusion: there is a strong correlation between physical movement of people and viral spread in the early stage of the \sars{} epidemic in Belgium, which weakens once the virus has spread nationally.\\

We first confirmed that mobility between geographical regions at NUTS 3 level (arrondissements) in Belgium was reduced during the first and second wave of the \covid{} pandemic in 2020. Second, we quantified time lag and morphological similarity between local \covid{}-related time series using dynamic time warping and time-lagged cross-correlation. This approach proved to be meaningful and intuitive, provided we focus on full-wave hospitalisation time series, and we consider the large standard deviation over results associated with noisy data. Third, we assessed the strength of the relationship between the connectivity index of pairs of arrondissements on the one hand, and DTW or TLCC lags on the other. Particularly, we quantified the strength of the relationship between the connectivity to arrondissements that are affected first on the one hand, and local excess deaths on the other hand. We demonstrated a \textit{strong} such correlation for the early-outbreak arrondissement Tongeren on the one hand, and a local excess mortality with a five to six week delay on the other hand.
More generally, we observed a significantly nonzero but \textit{weak} anticorrelation, notably for DTW distances. This confirms that, in our data set, strongly connected arrondissements exhibit morphologically similar hospitalisation time series, that are (on average) roughly synchronised. However, other factors beyond the control of our analysis appear to cloud a clean correlation.\\

The techniques developed in and conclusions drawn from this research demonstrate that a spatio-temporal data analysis of mobility and epidemiological data at NUTS 3 level in Belgium is feasible and informative. This motivates data analysis for other sociological and demographic aspects of society that may be employed as a metric for the \sars{} pandemic. Moreover, the conclusions imply that a model for \covid{} in Belgium may benefit from including a notion of mobility, especially when modelling the early stages of the \sars{} pandemic, before the virus has spread homogeneously throughout the country. We have therefore set up a spatially explicit model based on mobility data within Belgium \cite{rollier2022b}.

\backmatter

\paragraph{Supplementary information}

This article is associated with a supplementary document containing additional information on the Belgian geography, a more in-depth discussion of the GAMM fitting, and with additional results. It is currently included as an appendix.

\paragraph{Author contributions}

Michiel Rollier, Gisele Miranda, and Jenna Vergeynst contributed equally to this work. Jenna Vergeynst was responsible for early exploratory work with time series analysis, with early examination of mobility trends, and with the full analysis of excess deaths. Gisele Miranda was occupied with the realisation of GAMM fits, the calculation of corresponding DTW values and TLCC lags, and the conception of the heat plots with mean values and standard deviations of these quantities. Michiel Rollier was responsible for processing of the mobility data, development of the correlation coefficients for the scatter plots, and for general overview, formatting and wording of the document.

Joris Meys made important contributions regarding the technical background of GAMM fitting, as well as scrutinising statistical methods. Tijs Alleman contributed to the interpretation and processing of the \covid{} time series from a modeller's point of view.
The Belgian Collaborative Group on \covid{} processed and provided all data required for this analysis, and commented on its analysis.

Jan Baetens is Michiel Rollier's and Tijs Alleman's doctoral advisor. Both, and in particular the latter, gave formal advice and structural guidelines, and maintained the focus of this work.

All authors were in close communication and reviewed the final document. All authors consent to its submission to ``Mathematical Biosciences''.

\paragraph{Acknowledgements}

The authors wish to express their gratitude to Proximus, Belgium's leading telecom operator, for their generous commitment in the context of the Task Force `Data \& Technology against Corona'; in particular for providing daily detailed mobility data.

\paragraph{Conflict of interest} None declared.

\paragraph{Funding} This work was supported by the UGent \textit{Special Research Fund}, Belgium, by the \textit{Research Foundation Flanders (FWO)}, Belgium, project numbers G0G2920 and 3G0G9820 and by \textit{VZW 100 km Dodentocht Kadee}, Belgium through the organization of the 2020 100 km COVID-Challenge. Further, the computational resources and services used in this work were also provided by the VSC (\textit{Flemish Supercomputer Center}), funded by FWO, Belgium and the Flemish Government. The funding sources played no role in study design; in the collection, analysis and interpretation of data; in the writing of the report; nor in the decision to submit the article for publication.

\begin{appendices}

\section{Supplementary material}\label{sec:suppl-mat} 

\subsection{Additional geographical information}

\begin{figure}[h]
    \centering
    \includegraphics[width=.45\textwidth]{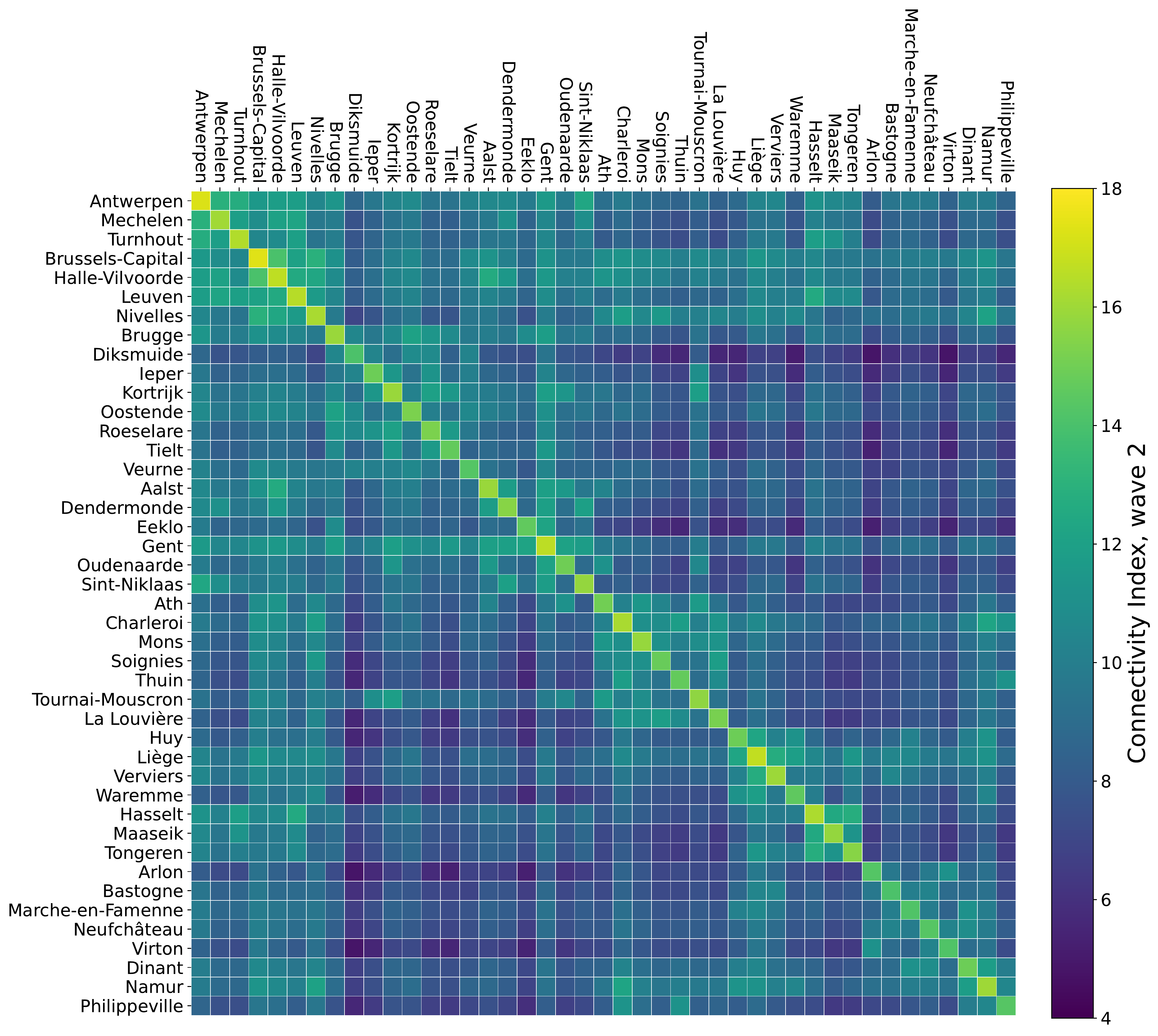} \hspace{.05\textwidth}
    \includegraphics[width=.45\textwidth]{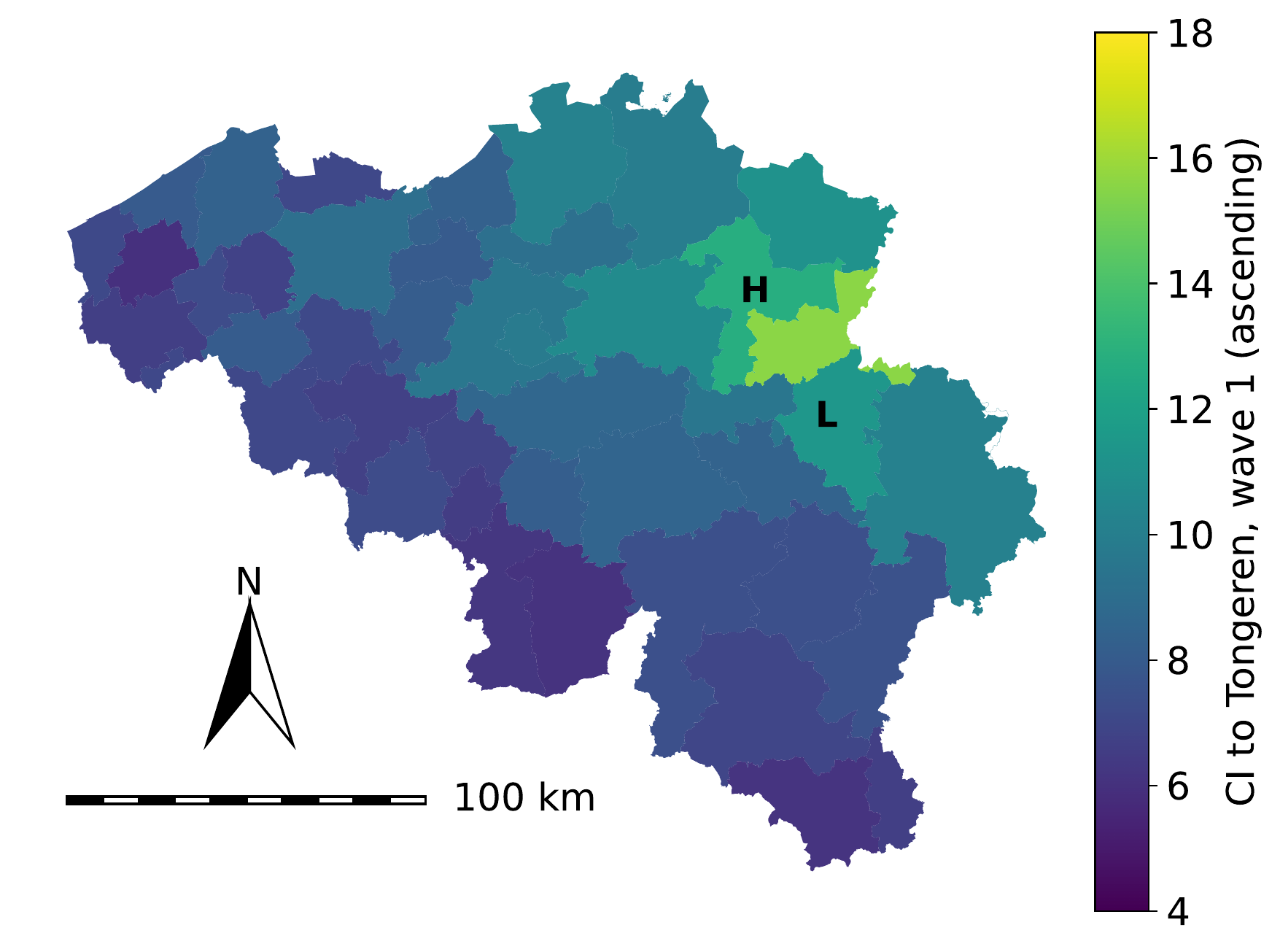}
    \caption{\textit{Left}: Heatmap showing the second-wave CIs, defined in Eq. \eqref{eq:CI-formula}. \textit{Right}: geographical detail of CIs to Tongeren during the ascending phase of the first wave, on the same colour scale, with an indication of arrondissements Hasselt and Liège.}
    \label{SI:fig:heatmap_wave2}
\end{figure}

Tab.~\ref{SI:tab:arr} contains all relevant demographic and geographical information on the 43 Belgian arrondissements. The left-hand side of Fig.~\ref{SI:fig:heatmap_wave2} demonstrates the symmetric matrix CI$^{gh}$ for all arrondissements for the \textit{second} \covid{} wave in Belgium; on the right we show the CI to Tongeren during the ascending phase of the first wave (compare this to Fig.~\ref{fig:CI-vs-ED}). Clearly, geographical distance is related to the CI.


\begin{table}
\footnotesize
\centering
\caption{43 Belgian arrondissements (NUTS 3) ordered by systematic identification (first two digits of NIS code) and blocked per province (NUTS 2), with population size, area and population density. Arrondissements containing a (province) capital are in boldface. The NUTS 1 regions are indicated by F (Flanders, Dutch names), W (Wallonia, French names) or B (Brussels-Capital Region, English name).}
\begin{tabular}{llrrrll}
\toprule
NIS & Arrondissement &  Pop. &  Area &  Density & Province & Region \\
& (NUTS 3) &  &   (km$^2$) &   (km$^{-2}$) & (NUTS 2) & (NUTS 1) \\

\midrule
11 & \textbf{Antwerpen}         &      1057736 &      1004 &     1053 &        Antwerpen & F \\
12 & Mechelen          &       347125 &       511 &      678 &        Antwerpen & F \\ \vspace{2pt}
13 & Turnhout          &       464869 &      1360 &      341 &        Antwerpen & F \\ \vspace{2pt}
21 & \textbf{Brussels-Capital} &      1218255 &       162 &     7500 &          N/A  & B \\
23 & Halle-Vilvoorde   &       643766 &       949 &      678 &   Vlaams-Brabant & F \\ \vspace{2pt}
24 & \textbf{Leuven}   &       512077 &      1169 &      437 &   Vlaams-Brabant & F \\ \vspace{2pt}
25 & Nivelles          &       406019 &      1097 &      370 &   Brabant Wallon & W \\
31 & \textbf{Brugge}   &       282745 &       673 &      419 &  West-Vlaanderen & F \\
32 & Diksmuide         &        51696 &       365 &      141 &  West-Vlaanderen & F \\
33 & Ieper             &       106570 &       553 &      192 &  West-Vlaanderen & F \\
34 & Kortrijk          &       292493 &       406 &      720 &  West-Vlaanderen & F \\
35 & Oostende          &       157780 &       304 &      518 &  West-Vlaanderen & F \\
36 & Roeselare         &       154494 &       273 &      564 &  West-Vlaanderen & F \\
37 & Tielt             &        93428 &       331 &      281 &  West-Vlaanderen & F \\ \vspace{2pt}
38 & Veurne            &        61739 &       288 &      214 &  West-Vlaanderen & F \\
41 & Aalst             &       293650 &       472 &      620 &  Oost-Vlaanderen & F \\
42 & Dendermonde       &       202411 &       346 &      584 &  Oost-Vlaanderen & F \\
43 & Eeklo             &        85692 &       335 &      255 &  Oost-Vlaanderen & F \\
44 & \textbf{Gent}     &       564042 &       949 &      593 &  Oost-Vlaanderen & F \\
45 & Oudenaarde        &       124610 &       422 &      294 &  Oost-Vlaanderen & F \\ \vspace{2pt}
46 & Sint-Niklaas      &       254850 &       479 &      531 &  Oost-Vlaanderen & F \\
51 & Ath               &       128468 &       671 &      191 &       Hainaut & W \\
52 & Charleroi         &       396962 &       475 &      834 &       Hainaut & W \\
53 & \textbf{Mons}     &       259237 &       588 &      440 &       Hainaut & W \\
55 & Soignies          &       105179 &       357 &      294 &       Hainaut & W \\
56 & Thuin             &        91725 &       785 &      116 &       Hainaut & W \\
57 & Tournai-Mouscron  &       223799 &       714 &      313 &       Hainaut & W \\ \vspace{2pt}
58 & La Louvière       &       141470 &       219 &      644 &       Hainaut & W \\
61 & Huy               &       113869 &       661 &      172 &             Li\`ege & W  \\
62 & \textbf{Li\`ege}  &       625765 &       795 &      786 &             Li\`ege & W  \\
63 & Verviers          &       288277 &      2009 &      143 &             Li\`ege & W  \\ \vspace{2pt}
64 & Waremme           &        81889 &       390 &      209 &             Li\`ege & W  \\
71 & \textbf{Hasselt}  &       420312 &       883 &      475 &          Limburg & F \\
72 & Maaseik           &       252115 &       910 &      276 &          Limburg & F  \\ \vspace{2pt}
73 & Tongeren          &       204943 &       633 &      323 &          Limburg & F  \\
81 & \textbf{Arlon}    &       62996 &       318 &      197 &        Luxembourg & W \\
82 & Bastogne          &        49083 &      1046 &       46 &        Luxembourg & W  \\
83 & Marche-en-Famenne &       56771 &       958 &       59 &        Luxembourg & W  \\
84 & Neufchâteau       &       63763 &      1358 &       46 &        Luxembourg & W  \\ \vspace{2pt}
85 & Virton            &       54139 &       777 &       69 &        Luxembourg & W  \\
91 & Dinant            &      111286 &      1596 &       69 &            Namur & W  \\
92 & \textbf{Namur}    &      318231 &      1167 &      272 &            Namur & W  \\ \vspace{2pt}
93 & Philippeville     &       66315 &       910 &       72 &            Namur & W  \\\bottomrule
\end{tabular}
\label{SI:tab:arr}
\end{table}

\begin{figure}[h]
    \centering
    \includegraphics[width=\textwidth]{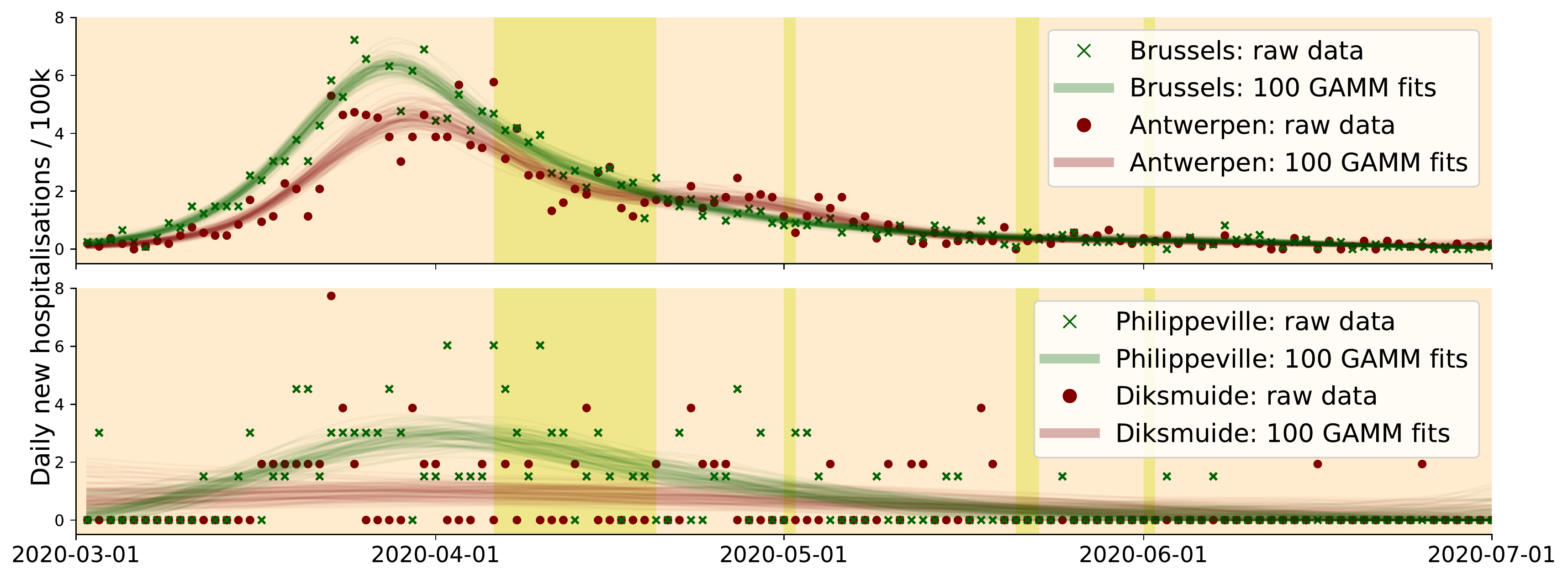}
    \caption{100 GAMM-fitted curves plotted over the original data for daily new \covid{} hospitalisations during the first 2020 epidemic wave in the arrondissements Brussels-Capital, Antwerpen, Philippeville and Diksmuide.}
    \label{fig:GAMM-fits-examples}
\end{figure}

\pagebreak
\subsection{Details of the time series GAMM fitting}

\paragraph{Mathematical construction}

Our approach for data smoothing and uncertainty indication makes use of GAMM fitting and bootstrapping \citep{Wood2017}, described in the main text. We carried out all calculations with the \texttt{mgcv} package \citep{Wood2017} in the statistical software R \citep{Rsoftware2021}. Note, for completeness, that due to the log transformation the model fit becomes unstable if there are long periods without events. To increase numerical stability, we simply add 1 to the data prior to fitting in these cases. This correction is subsequently subtracted from the resulting predictions.

\paragraph{Assumptions and weaknesses}

We use GAMMs as a more advanced smoothing technique and as a means to indicate uncertainty. Contrary to moving averages and other local smoothing techniques, they allow us to incorporate both autocorrelation and overdispersion. The resulting smooth curves are in general less sensitive to erratic fluctuations in the data, while still sufficiently flexible to describe general trends (and differences between these) in a time series of number of events \citep{eilers1996,Wood2017}.\\

The GAMM framework also offers a more formal estimation of the uncertainty on the parameters. This allows for a computationally efficient method to construct bootstrap samples from these general trends, which are in turn used to formulate a spectrum of slightly deviating results. A possible weakness lies in the fact that this estimate of uncertainty relies heavily on a number of assumptions. First of all, we assume that the $\bm{\beta}$ coefficients follow a multivariate normal distribution. While this distribution is not guaranteed, the Laplace approximation performs well for a sufficiently large sample size \citep{kauermann2009}. We assume that the amount of data used is sufficient to expect little deviation from this assumption. \\

Second, we assume a linear relationship between the mean and variance of predictions. This approach, often referred to as quasi-Poisson, has been used in numerous other analyses (e.g. Refs.~\cite{ouldali2020,angoulvant2020,vicuna2021}). Yet, assuming a quadratic relationship between mean and variance would be more equivalent to the negative binomial distribution assumed by Endo et al.~\cite{endo2020}. On the other hand, such an approach would give larger values less weight in the fit compared to the quasi-Poisson method \citep{VerHoef2007}. Comparison of both methods for a selection of arrondissements showed that assuming a quadratic relationship would lead to a systematic underestimation of the peak height. \\ 

By analysing the squared deviation from weekly averages, we concluded that the linear assumption could be defended. Only for the few cases with a very large number of events, variance would be underestimated. In the majority of cases, the linear relation would slightly overestimate the variance, making the approach more conservative.





\subsection{Additional DTW and TLCC results}

\begin{figure}[h]
    \centering
    \includegraphics[width=0.9\textwidth]{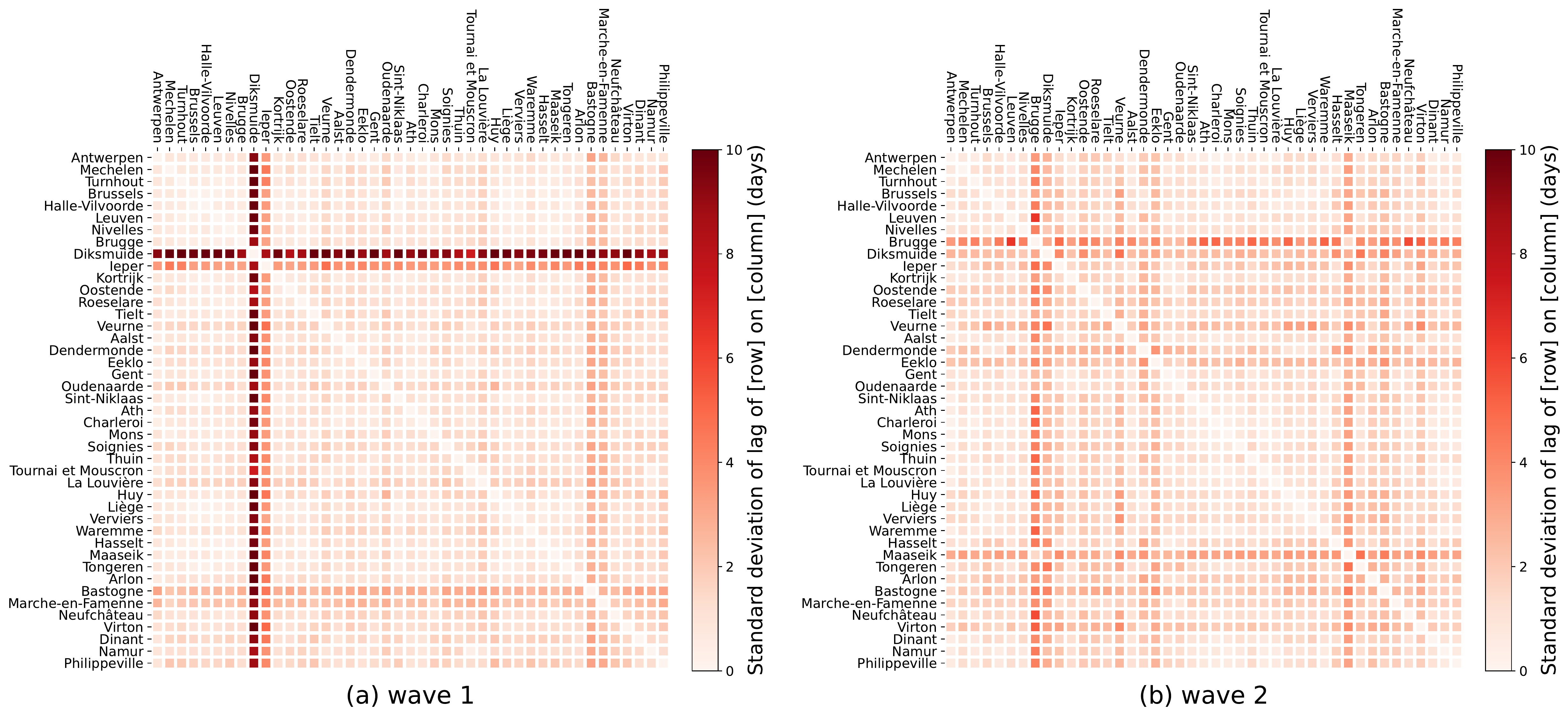}\\
    \includegraphics[width=0.9\textwidth]{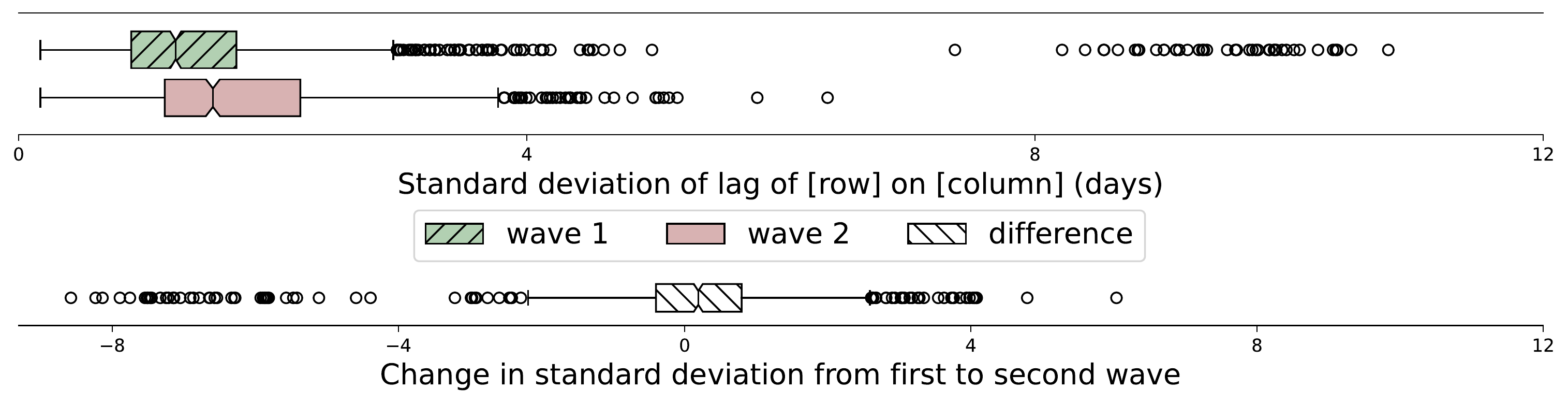}
    \caption{Formatting as in Fig.~\ref{fig:TLCC_hosp}. \textit{Top:} Standard deviation of all 100 normalised dynamic time warping distances computed between each unique pair of Belgian arrondissements from 100 GAMM fits on the hospitalisation time series, for full first (a) and second (b) \covid{} waves. \textit{Bottom:} Boxplots for the same 903 unique arrondissement pairs, demonstrating visually the difference in distributions (Wilcoxon test $p$ value $\sim 10^{-7}$).}
    \label{fig:TLCC_hosp_std}
\end{figure}

\begin{figure}[h]
    \centering
    \includegraphics[width=0.9\textwidth]{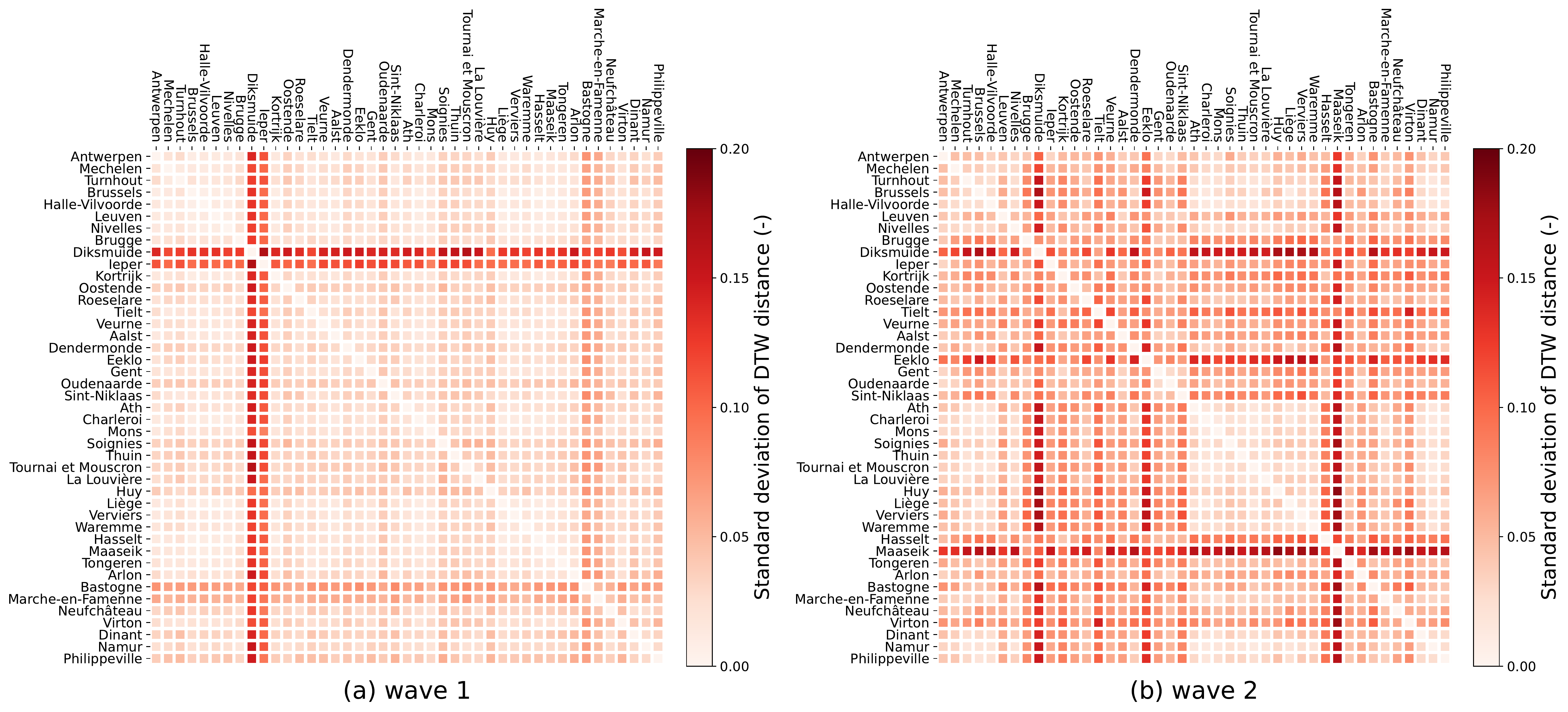}\\
    \includegraphics[width=0.9\textwidth]{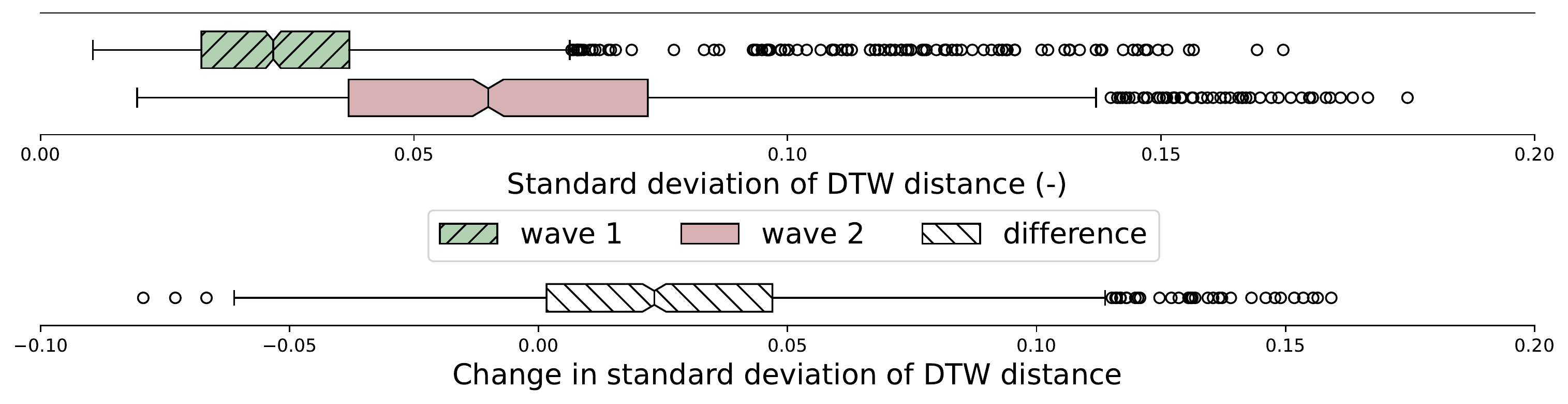}
    \caption{Formatting as in Fig.~\ref{fig:DTW_hosp}, again communicating standard deviations as in Fig.~\ref{fig:TLCC_hosp_std}, but now for DTW values of arrondissement pairs. Wilcoxon test $p$ value for differences $\sim 10^{-73}$.}
    \label{fig:DTW_hosp_std}
\end{figure}

The heatmaps and boxplots in Fig.~\ref{fig:TLCC_hosp_std}, associated with Fig.~\ref{fig:TLCC_hosp}, communicate the standard deviations of TLCC lags, taken over 100 slightly different GAMM realisations of the hospitalisation time series of 903 unique arrondissement pairs, for the entire first \covid{} wave in Belgium. Fig.~\ref{fig:DTW_hosp_std}, associated with Fig.~\ref{fig:DTW_hosp}, shows the same, but for DTW values. A comprehensive table for all results is provided in Tab.~\ref{tab:additional_correlation_results}.

\begin{table}[]
    \centering
    \caption{Additional Spearman's rank correlation coefficients for the correlation between DTW values and TLCC lags on the one hand, and the connectivity index on the other. Mean values and standard deviations of 100 GAMM fits are provided for daily new \covid{} hospitalisation time series at the arrondissement level, for both 2020 waves.}
    \begin{tabular}{lrrrr}
        & \multicolumn{4}{c}{Spearman's rank correlation coefficient} \\ \toprule
         & \multicolumn{2}{c}{Wave 1} & \multicolumn{2}{c}{Wave 2} \\ \midrule
         & \multicolumn{1}{c}{TLCC} & \multicolumn{1}{c}{DTW} & \multicolumn{1}{c}{TLCC} & \multicolumn{1}{c}{DTW} \\
        \ Full & -0.16(4) & -0.25(4) & -0.15(4) & -0.31(3) \\
        \ Asc. & -0.14(4) & -0.15(5) & -0.11(5) & -0.16(4)\\
        \ Desc. & -0.14(4) & -0.19(3) & -0.11(6) & -0.31(4)\\ \bottomrule
    \end{tabular}
    \label{tab:additional_correlation_results}
\end{table}

\end{appendices}


{\footnotesize
\bibliography{bibliography}}


\begin{thebibliography}{44}
\ifx \bisbn   \undefined \def \bisbn  #1{ISBN #1}\fi
\ifx \binits  \undefined \def \binits#1{#1}\fi
\ifx \bauthor  \undefined \def \bauthor#1{#1}\fi
\ifx \batitle  \undefined \def \batitle#1{#1}\fi
\ifx \bjtitle  \undefined \def \bjtitle#1{#1}\fi
\ifx \bvolume  \undefined \def \bvolume#1{\textbf{#1}}\fi
\ifx \byear  \undefined \def \byear#1{#1}\fi
\ifx \bissue  \undefined \def \bissue#1{#1}\fi
\ifx \bfpage  \undefined \def \bfpage#1{#1}\fi
\ifx \blpage  \undefined \def \blpage #1{#1}\fi
\ifx \burl  \undefined \def \burl#1{\textsf{#1}}\fi
\ifx \doiurl  \undefined \def \doiurl#1{\url{https://doi.org/#1}}\fi
\ifx \betal  \undefined \def \betal{\textit{et al.}}\fi
\ifx \binstitute  \undefined \def \binstitute#1{#1}\fi
\ifx \binstitutionaled  \undefined \def \binstitutionaled#1{#1}\fi
\ifx \bctitle  \undefined \def \bctitle#1{#1}\fi
\ifx \beditor  \undefined \def \beditor#1{#1}\fi
\ifx \bpublisher  \undefined \def \bpublisher#1{#1}\fi
\ifx \bbtitle  \undefined \def \bbtitle#1{#1}\fi
\ifx \bedition  \undefined \def \bedition#1{#1}\fi
\ifx \bseriesno  \undefined \def \bseriesno#1{#1}\fi
\ifx \blocation  \undefined \def \blocation#1{#1}\fi
\ifx \bsertitle  \undefined \def \bsertitle#1{#1}\fi
\ifx \bsnm \undefined \def \bsnm#1{#1}\fi
\ifx \bsuffix \undefined \def \bsuffix#1{#1}\fi
\ifx \bparticle \undefined \def \bparticle#1{#1}\fi
\ifx \barticle \undefined \def \barticle#1{#1}\fi
\bibcommenthead
\ifx \bconfdate \undefined \def \bconfdate #1{#1}\fi
\ifx \botherref \undefined \def \botherref #1{#1}\fi
\ifx \url \undefined \def \url#1{\textsf{#1}}\fi
\ifx \bchapter \undefined \def \bchapter#1{#1}\fi
\ifx \bbook \undefined \def \bbook#1{#1}\fi
\ifx \bcomment \undefined \def \bcomment#1{#1}\fi
\ifx \oauthor \undefined \def \oauthor#1{#1}\fi
\ifx \citeauthoryear \undefined \def \citeauthoryear#1{#1}\fi
\ifx \endbibitem  \undefined \def \endbibitem {}\fi
\ifx \bconflocation  \undefined \def \bconflocation#1{#1}\fi
\ifx \arxivurl  \undefined \def \arxivurl#1{\textsf{#1}}\fi
\csname PreBibitemsHook\endcsname

\bibitem{FPS1}
\begin{botherref}
\oauthor{\bsnm{Health}, \binits{F.P.S.}}:
{One repatriated Belgian has tested positive for the novel coronavirus}
(2020).
\url{https://web.archive.org/web/20200406101412/https://www.info-coronavirus.be/en/news/one-repatriated-belgian-has-tested-positive-for-the-novel-coronavirus/}
Accessed 2020-02-04
\end{botherref}
\endbibitem

\bibitem{FPS2}
\begin{botherref}
\oauthor{\bsnm{Health}, \binits{F.P.S.}}:
{Returns from Italy push COVID-19 tally higher}
(2020).
\url{https://www.vrt.be/vrtnws/en/2020/03/02/returns-from-italy-push-covid-19-tally-higher/}
Accessed 2020-03-02
\end{botherref}
\endbibitem

\bibitem{Sciensano2020}
\begin{botherref}
\oauthor{\bsnm{Sciensano}}:
{EPISTAT}.
Available at
  \url{https://statbel.fgov.be/en/open-data/number-deaths-day-sex-district-age}
(2020).
\url{https://epistat.wiv-isp.be/covid/}
\end{botherref}
\endbibitem

\bibitem{Alleman2021}
\begin{barticle}
\bauthor{\bsnm{Alleman}, \binits{T.W.}},
\bauthor{\bsnm{Vergeynst}, \binits{J.}},
\bauthor{\bsnm{Visscher}, \binits{L.D.}},
\bauthor{\bsnm{Rollier}, \binits{M.}},
\bauthor{\bsnm{Torfs}, \binits{E.}},
\bauthor{\bsnm{{the Belgian Collaborative Group on COVID-19 Hospital
  Surveillance}}},
\bauthor{\bsnm{Nopens}, \binits{I.}},
\bauthor{\bsnm{Baetens}, \binits{J.M.}}:
\batitle{{Assessing the effects of non-pharmaceutical interventions on
  SARS-CoV-2 spread in Belgium by means of a compartmental , age-stratified ,
  extended SEIQRD model and public mobility data}}.
\bjtitle{Epidemics}
\bvolume{37},
\bfpage{100505}
(\byear{2021})
\end{barticle}
\endbibitem

\bibitem{Abrams2020}
\begin{barticle}
\bauthor{\bsnm{Abrams}, \binits{S.}},
\bauthor{\bsnm{Wambua}, \binits{J.}},
\bauthor{\bsnm{Santermans}, \binits{E.}},
\bauthor{\bsnm{Willem}, \binits{L.}},
\bauthor{\bsnm{Kuylen}, \binits{E.}},
\bauthor{\bsnm{Coletti}, \binits{P.}},
\bauthor{\bsnm{Libin}, \binits{P.}},
\bauthor{\bsnm{Faes}, \binits{C.}},
\bauthor{\bsnm{Petrof}, \binits{O.}},
\bauthor{\bsnm{Herzog}, \binits{S.A.}},
\bauthor{\bsnm{Beutels}, \binits{P.}},
\bauthor{\bsnm{Hens}, \binits{N.}}:
\batitle{Modelling the early phase of the belgian covid-19 epidemic using a
  stochastic compartmental model and studying its implied future trajectories}.
\bjtitle{Epidemics}
\bvolume{35},
\bfpage{100449}
(\byear{2021}).
\doiurl{10.1016/j.epidem.2021.100449}
\end{barticle}
\endbibitem

\bibitem{Franco2020}
\begin{barticle}
\bauthor{\bsnm{Franco}, \binits{N.}}:
\batitle{Covid-19 belgium: Extended seir-qd model with nursing homes and
  long-term scenarios-based forecasts}.
\bjtitle{Epidemics}
\bvolume{37},
\bfpage{100490}
(\byear{2021}).
\doiurl{10.1016/j.epidem.2021.100490}
\end{barticle}
\endbibitem

\bibitem{RESTORE8}
\begin{botherref}
\oauthor{\bsnm{\textsc{restore}}}:
RESTORE Report 8: Long-term scenarios for the number of new hospitalizations
  during the Belgian COVID-19 epidemic
(2021).
\url{https://covid-en-wetenschap.github.io/assets/restore/report_v8_0.pdf}
Accessed 2021-06-29
\end{botherref}
\endbibitem

\bibitem{Changruenngam2020}
\begin{barticle}
\bauthor{\bsnm{Changruenngam}, \binits{S.}},
\bauthor{\bsnm{Bicout}, \binits{D.J.}},
\bauthor{\bsnm{Modchang}, \binits{C.}}:
\batitle{{How the individual human mobility spatio-temporally shapes the
  disease transmission dynamics}}.
\bjtitle{Scientific Reports}
\bvolume{10}(\bissue{1}),
\bfpage{1}--\blpage{13}
(\byear{2020}).
\doiurl{10.1038/s41598-020-68230-9}
\end{barticle}
\endbibitem

\bibitem{Merler2010}
\begin{barticle}
\bauthor{\bsnm{Merler}, \binits{S.}},
\bauthor{\bsnm{Ajelli}, \binits{M.}}:
\batitle{{The role of population heterogeneity and human mobility in the spread
  of pandemic influenza}}.
\bjtitle{Proceedings of the Royal Society B: Biological Sciences}
\bvolume{277}(\bissue{1681}),
\bfpage{557}--\blpage{565}
(\byear{2010}).
\doiurl{10.1098/rspb.2009.1605}
\end{barticle}
\endbibitem

\bibitem{Wesolowski2012}
\begin{barticle}
\bauthor{\bsnm{Wesolowski}, \binits{A.}},
\bauthor{\bsnm{Eagle}, \binits{N.}},
\bauthor{\bsnm{Tatem}, \binits{A.J.}},
\bauthor{\bsnm{Smith}, \binits{D.L.}},
\bauthor{\bsnm{Noor}, \binits{A.M.}},
\bauthor{\bsnm{Snow}, \binits{R.W.}},
\bauthor{\bsnm{Buckee}, \binits{C.O.}}:
\batitle{{Quantifying the impact of human mobility on malaria}}.
\bjtitle{Science}
\bvolume{338}(\bissue{6104}),
\bfpage{267}--\blpage{270}
(\byear{2012}).
\doiurl{10.1126/science.1223467}
\end{barticle}
\endbibitem

\bibitem{Wesolowski2015}
\begin{barticle}
\bauthor{\bsnm{Wesolowski}, \binits{A.}},
\bauthor{\bsnm{Qureshi}, \binits{T.}},
\bauthor{\bsnm{Boni}, \binits{M.F.}},
\bauthor{\bsnm{Sunds{\o}y}, \binits{P.R.}},
\bauthor{\bsnm{Johansson}, \binits{M.A.}},
\bauthor{\bsnm{Rasheed}, \binits{S.B.}},
\bauthor{\bsnm{Eng{\o}-Monsen}, \binits{K.}},
\bauthor{\bsnm{Buckee}, \binits{C.O.}}:
\batitle{{Impact of human mobility on the emergence of dengue epidemics in
  Pakistan}}.
\bjtitle{Proceedings of the National Academy of Sciences of the United States
  of America}
\bvolume{112}(\bissue{38}),
\bfpage{11887}--\blpage{11892}
(\byear{2015}).
\doiurl{10.1073/pnas.1504964112}
\end{barticle}
\endbibitem

\bibitem{du2020}
\begin{barticle}
\bauthor{\bsnm{Du}, \binits{Z.}},
\bauthor{\bsnm{Wang}, \binits{L.}},
\bauthor{\bsnm{Cauchemez}, \binits{S.}},
\bauthor{\bsnm{Xu}, \binits{X.}},
\bauthor{\bsnm{Wang}, \binits{X.}},
\bauthor{\bsnm{Cowling}, \binits{B.J.}},
\bauthor{\bsnm{Meyee}, \binits{L.A.}}:
\batitle{Transportation of coronavirus disease from wuhan to other cities in
  china}.
\bjtitle{Emerging Infectious Diseases}
\bvolume{26}(\bissue{5}),
\bfpage{1049}--\blpage{1052}
(\byear{2020}).
\doiurl{10.3201/eid2605.200146}
\end{barticle}
\endbibitem

\bibitem{jia2020}
\begin{barticle}
\bauthor{\bsnm{Jia}, \binits{J.}},
\bauthor{\bsnm{Lu}, \binits{X.}},
\bauthor{\bsnm{Yuan}, \binits{Y.}},
\bauthor{\bsnm{Xu}, \binits{G.}},
\bauthor{\bsnm{Jia}, \binits{J.}},
\bauthor{\bsnm{Christakis}, \binits{N.}}:
\batitle{Population flow drives spatio-temporal distribution of covid-19 in
  china}.
\bjtitle{Nature}
(\byear{2020}).
\doiurl{10.1038/s41586-020-2284-y}
\end{barticle}
\endbibitem

\bibitem{Iacus2020}
\begin{botherref}
\oauthor{\bsnm{Iacus}, \binits{S.M.}},
\oauthor{\bsnm{Santamaria}, \binits{C.}},
\oauthor{\bsnm{Sermi}, \binits{F.}},
\oauthor{\bsnm{Spyratos}, \binits{S.}},
\oauthor{\bsnm{Tarchi}, \binits{D.}},
\oauthor{\bsnm{{Vespe M.}}}:
{How human mobility explains the initial spread of COVID-19}.
Technical report,
Joint Research Centre
(2020).
\doiurl{10.2760/61847}
\end{botherref}
\endbibitem

\bibitem{Arenas2020}
\begin{botherref}
\oauthor{\bsnm{Arenas}, \binits{A.}},
\oauthor{\bsnm{Cota}, \binits{W.}},
\oauthor{\bsnm{Gomez-Gardenes}, \binits{J.}},
\oauthor{\bsnm{G{\'{o}}mez}, \binits{S.}},
\oauthor{\bsnm{Granell}, \binits{C.}},
\oauthor{\bsnm{Matamalas}, \binits{J.T.}},
\oauthor{\bsnm{Soriano-Panos}, \binits{D.}},
\oauthor{\bsnm{Steinegger}, \binits{B.}}:
{A mathematical model for the spatiotemporal epidemic spreading of COVID19}.
medRxiv,
2020--032120040022
(2020).
\doiurl{10.1101/2020.03.21.20040022}
\end{botherref}
\endbibitem

\bibitem{Costa2020}
\begin{botherref}
\oauthor{\bsnm{Costa}, \binits{G.S.}},
\oauthor{\bsnm{Cota}, \binits{W.}},
\oauthor{\bsnm{Ferreira}, \binits{S.C.}}:
{Metapopulation modeling of COVID-19 advancing into the countryside: an
  analysis of mitigation strategies for Brazil}.
medRxiv,
2020--050620093492
(2020).
\doiurl{10.1101/2020.05.06.20093492}
\end{botherref}
\endbibitem

\bibitem{Roques2020}
\begin{barticle}
\bauthor{\bsnm{Roques}, \binits{L.}},
\bauthor{\bsnm{Bonnefon}, \binits{O.}},
\bauthor{\bsnm{Baudrot}, \binits{V.}},
\bauthor{\bsnm{Soubeyrand}, \binits{S.}},
\bauthor{\bsnm{Berestycki}, \binits{H.}}:
\batitle{{A parsimonious approach for spatial transmission and heterogeneity in
  the COVID-19 propagation: Modelling the COVID-19 propagation}}.
\bjtitle{Royal Society Open Science}
\bvolume{7}(\bissue{12}),
\bfpage{1}--\blpage{19}
(\byear{2020}).
\doiurl{10.1098/rsos.201382}
\end{barticle}
\endbibitem

\bibitem{sartorius2021}
\begin{barticle}
\bauthor{\bsnm{Sartorius}, \binits{B.}},
\bauthor{\bsnm{Lawson}, \binits{A.}},
\bauthor{\bsnm{Pullan}, \binits{R.}}:
\batitle{Modelling and predicting the spatio-temporal spread of covid-19,
  associated deaths and impact of key risk factors in england.}
\bjtitle{Scientific reports}
\bvolume{11}(\bissue{1}),
\bfpage{5378}
(\byear{2021})
\end{barticle}
\endbibitem

\bibitem{Knoema2011}
\begin{botherref}
\oauthor{\bsnm{Knoema}}:
{Belgium - Road density}
(2011).
\url{https://knoema.com/atlas/Belgium/Road-density}
\end{botherref}
\endbibitem

\bibitem{STATBEL2020}
\begin{botherref}
\oauthor{\bsnm{Statbel}}:
{Bevolkingsdichtheid}.
Available at
  \url{https://statbel.fgov.be/nl/themas/bevolking/bevolkingsdichtheid}
(2020).
\url{https://statbel.fgov.be/nl/themas/bevolking/bevolkingsdichtheid}
\end{botherref}
\endbibitem

\bibitem{rollier2022b}
\begin{botherref}
\oauthor{\bsnm{Rollier}, \binits{M.}},
\oauthor{\bsnm{Alleman}, \binits{T.}},
\oauthor{\bsnm{Vergeynst}, \binits{J.}},
\oauthor{\bsnm{Baetens}, \binits{J.M.}}:
A Mobility-Driven Spatially Explicit SEIQRD COVID-19 Model with VOCs,
  seasonality, and vaccines.
arXiv
(2022).
\doiurl{10.48550/ARXIV.2207.03717}.
\url{https://arxiv.org/abs/2207.03717}
\end{botherref}
\endbibitem

\bibitem{Statbel2020a}
\begin{botherref}
\oauthor{\bsnm{Statbel}}:
{Number of deaths per day, sex, age, region, province, district}.
Available at
  \url{https://statbel.fgov.be/en/open-data/number-deaths-day-sex-district-age}
(2020).
\url{https://statbel.fgov.be/en/open-data/number-deaths-day-sex-district-age}
\end{botherref}
\endbibitem

\bibitem{karanikolos2020}
\begin{botherref}
\oauthor{\bsnm{Karanikolos}, \binits{M.}},
\oauthor{\bsnm{McKee}, \binits{M.}}:
{How Comparable Is COVID-19 Mortality Across Countries?}
\url{https://analysis.covid19healthsystem.org/index.php/2020/06/04/how-comparable-is-covid-19-mortality-across-countries/}.
Accessed: 2022-06-15
(2020)
\end{botherref}
\endbibitem

\bibitem{FOD_economie_proximus_market-share}
\begin{botherref}
\oauthor{\bsnm{{Federal Public Service Economy}}}:
{Belgische telecommunicatie- en televisiesectoren}
(2019).
\url{https://economie.fgov.be/nl/themas/online/telecommunicatie/belgische-telecommunicatie-en}
Accessed 2021-05-18
\end{botherref}
\endbibitem

\bibitem{Wood2017}
\begin{bbook}
\bauthor{\bsnm{Wood}, \binits{S.N.}}:
\bbtitle{Generalized Additive Models: An Introduction with R, Second Edition}.
\bsertitle{Chapman \& Hall/CRC Texts in Statistical Science}.
\bpublisher{CRC Press}, \blocation{???}
(\byear{2017})
\end{bbook}
\endbibitem

\bibitem{eilers1996}
\begin{barticle}
\bauthor{\bsnm{Eilers}, \binits{P.H.C.}},
\bauthor{\bsnm{Marx}, \binits{B.D.}}:
\batitle{{Flexible smoothing with B-splines and penalties}}.
\bjtitle{Statistical Science}
\bvolume{11}(\bissue{2}),
\bfpage{89}--\blpage{121}
(\byear{1996}).
\doiurl{10.1214/ss/1038425655}
\end{barticle}
\endbibitem

\bibitem{Scortichini2020}
\begin{barticle}
\bauthor{\bsnm{Scortichini}, \binits{M.}},
\bauthor{\bparticle{Schneider~dos} \bsnm{Santos}, \binits{R.}},
\bauthor{\bsnm{De’~Donato}, \binits{F.}},
\bauthor{\bsnm{De~Sario}, \binits{M.}},
\bauthor{\bsnm{Michelozzi}, \binits{P.}},
\bauthor{\bsnm{Davoli}, \binits{M.}},
\bauthor{\bsnm{Masselot}, \binits{P.}},
\bauthor{\bsnm{Sera}, \binits{F.}},
\bauthor{\bsnm{Gasparrini}, \binits{A.}}:
\batitle{{Excess mortality during the COVID-19 outbreak in Italy: a two-stage
  interrupted time-series analysis}}.
\bjtitle{International Journal of Epidemiology}
\bvolume{49}(\bissue{6}),
\bfpage{1909}--\blpage{1917}
(\byear{2020})
{\href{https://arxiv.org/abs/https://academic.oup.com/ije/article-pdf/49/6/1909/36083908/dyaa169.pdf}{{https://academic.oup.com/ije/article-pdf/49/6/1909/36083908/dyaa169.pdf}}}.
\doiurl{10.1093/ije/dyaa169}
\end{barticle}
\endbibitem

\bibitem{shen2015analysis}
\begin{barticle}
\bauthor{\bsnm{Shen}, \binits{C.}}:
\batitle{Analysis of detrended time-lagged cross-correlation between two
  nonstationary time series}.
\bjtitle{Physics Letters A}
\bvolume{379}(\bissue{7}),
\bfpage{680}--\blpage{687}
(\byear{2015})
\end{barticle}
\endbibitem

\bibitem{paploski2016time}
\begin{barticle}
\bauthor{\bsnm{Paploski}, \binits{I.A.}},
\bauthor{\bsnm{Prates}, \binits{A.P.P.}},
\bauthor{\bsnm{Cardoso}, \binits{C.W.}},
\bauthor{\bsnm{Kikuti}, \binits{M.}},
\bauthor{\bsnm{Silva}, \binits{M.M.}},
\bauthor{\bsnm{Waller}, \binits{L.A.}},
\bauthor{\bsnm{Reis}, \binits{M.G.}},
\bauthor{\bsnm{Kitron}, \binits{U.}},
\bauthor{\bsnm{Ribeiro}, \binits{G.S.}}:
\batitle{Time lags between exanthematous illness attributed to zika virus,
  guillain-barr{\'e} syndrome, and microcephaly, salvador, brazil}.
\bjtitle{Emerging infectious diseases}
\bvolume{22}(\bissue{8}),
\bfpage{1438}
(\byear{2016})
\end{barticle}
\endbibitem

\bibitem{Kongming1997}
\begin{barticle}
\bauthor{\bsnm{Wang}, \binits{K.}},
\bauthor{\bsnm{Gasser}, \binits{T.}}:
\batitle{{Alignment of curves by dynamic time warping}}.
\bjtitle{The Annals of Statistics}
\bvolume{25}(\bissue{3}),
\bfpage{1251}--\blpage{1276}
(\byear{1997}).
\doiurl{10.1214/aos/1069362747}
\end{barticle}
\endbibitem

\bibitem{giorgino2009computing}
\begin{barticle}
\bauthor{\bsnm{Giorgino}, \binits{T.}}, \betal:
\batitle{Computing and visualizing dynamic time warping alignments in r: the
  dtw package}.
\bjtitle{Journal of statistical Software}
\bvolume{31}(\bissue{7}),
\bfpage{1}--\blpage{24}
(\byear{2009})
\end{barticle}
\endbibitem

\bibitem{spearman1987proof}
\begin{barticle}
\bauthor{\bsnm{Spearman}, \binits{C.}}:
\batitle{The proof and measurement of association between two things}.
\bjtitle{The American journal of psychology}
\bvolume{100}(\bissue{3/4}),
\bfpage{441}--\blpage{471}
(\byear{1987})
\end{barticle}
\endbibitem

\bibitem{vanmeeteren2016}
\begin{botherref}
\oauthor{\bsnm{{van Meeteren}}, \binits{M.}},
\oauthor{\bsnm{Bossauw}, \binits{K.}}:
{Metropoolvorming in België en Vlaanderen:De polycentrische ruimtelijke
  structuur van de arbeidsmarkt}.
Technical report,
Steunpunt Ruimte
(2016).
\doiurl{10.13140/RG.2.1.3718.4248}
\end{botherref}
\endbibitem

\bibitem{Islam2021}
\begin{botherref}
\oauthor{\bsnm{Islam}, \binits{N.}},
\oauthor{\bsnm{Shkolnikov}, \binits{V.M.}},
\oauthor{\bsnm{Acosta}, \binits{R.J.}},
\oauthor{\bsnm{Klimkin}, \binits{I.}},
\oauthor{\bsnm{Kawachi}, \binits{I.}},
\oauthor{\bsnm{Irizarry}, \binits{R.A.}},
\oauthor{\bsnm{Alicandro}, \binits{G.}},
\oauthor{\bsnm{Khunti}, \binits{K.}},
\oauthor{\bsnm{Yates}, \binits{T.}},
\oauthor{\bsnm{Jdanov}, \binits{D.A.}},
\oauthor{\bsnm{White}, \binits{M.}},
\oauthor{\bsnm{Lewington}, \binits{S.}},
\oauthor{\bsnm{Lacey}, \binits{B.}}:
{Excess deaths associated with covid-19 pandemic in 2020: age and sex
  disaggregated time series analysis in 29 high income countries}.
BMJ
\textbf{373}
(2021).
\doiurl{10.1136/BMJ.N1137}
\end{botherref}
\endbibitem

\bibitem{Wilcoxon1945}
\begin{barticle}
\bauthor{\bsnm{Wilcoxon}, \binits{F.}}:
\batitle{Individual comparisons by ranking methods}.
\bjtitle{Biometrics Bulletin}
\bvolume{1}(\bissue{6}),
\bfpage{80}--\blpage{83}
(\byear{1945}).
Accessed 2022-11-09
\end{barticle}
\endbibitem

\bibitem{Lauer2020}
\begin{barticle}
\bauthor{\bsnm{Lauer}, \binits{S.A.}},
\bauthor{\bsnm{Grantz}, \binits{K.H.}},
\bauthor{\bsnm{Bi}, \binits{Q.}},
\bauthor{\bsnm{Jones}, \binits{F.K.}},
\bauthor{\bsnm{Zheng}, \binits{Q.}},
\bauthor{\bsnm{Meredith}, \binits{H.R.}},
\bauthor{\bsnm{Azman}, \binits{A.S.}},
\bauthor{\bsnm{Reich}, \binits{N.G.}},
\bauthor{\bsnm{Lessler}, \binits{J.}}:
\batitle{{The incubation period of coronavirus disease 2019 (CoVID-19) from
  publicly reported confirmed cases: Estimation and application}}.
\bjtitle{Annals of Internal Medicine}
\bvolume{172}(\bissue{9}),
\bfpage{577}--\blpage{582}
(\byear{2020}).
\doiurl{10.7326/M20-0504}
\end{barticle}
\endbibitem

\bibitem{Habib2021}
\begin{barticle}
\bauthor{\bsnm{Habib}, \binits{Y.}},
\bauthor{\bsnm{Xia}, \binits{E.}},
\bauthor{\bsnm{Hashmi}, \binits{S.H.}},
\bauthor{\bsnm{Fareed}, \binits{Z.}}:
\batitle{{Non-linear spatial linkage between COVID-19 pandemic and mobility in
  ten countries: A lesson for future wave}}.
\bjtitle{Journal of Infection and Public Health}
\bvolume{14}(\bissue{10}),
\bfpage{1411}--\blpage{1426}
(\byear{2021}).
\doiurl{10.1016/J.JIPH.2021.08.008}
\end{barticle}
\endbibitem

\bibitem{Rsoftware2021}
\begin{bbook}
\bauthor{\bsnm{{R Core Team}}}:
\bbtitle{R: A Language and Environment for Statistical Computing}.
\bpublisher{R Foundation for Statistical Computing},
\blocation{Vienna, Austria}
(\byear{2021}).
\bcomment{R Foundation for Statistical Computing}.
\burl{https://www.R-project.org/}
\end{bbook}
\endbibitem

\bibitem{kauermann2009}
\begin{barticle}
\bauthor{\bsnm{Kauermann}, \binits{G.}},
\bauthor{\bsnm{Krivobokova}, \binits{T.}},
\bauthor{\bsnm{Fahrmeir}, \binits{L.}}:
\batitle{Some asymptotic results on generalized penalized spline smoothing}.
\bjtitle{Journal of the Royal Statistical Society: Series B (Statistical
  Methodology)}
\bvolume{71}(\bissue{2}),
\bfpage{487}--\blpage{503}
(\byear{2009}).
\doiurl{10.1111/j.1467-9868.2008.00691.x}
\end{barticle}
\endbibitem

\bibitem{ouldali2020}
\begin{barticle}
\bauthor{\bsnm{Ouldali}, \binits{N.}},
\bauthor{\bsnm{Pouletty}, \binits{M.}},
\bauthor{\bsnm{Mariani}, \binits{P.}},
\bauthor{\bsnm{Beyler}, \binits{C.}},
\bauthor{\bsnm{Blachier}, \binits{A.}},
\bauthor{\bsnm{Bonacorsi}, \binits{S.}},
\bauthor{\bsnm{Danis}, \binits{K.}},
\bauthor{\bsnm{Chomton}, \binits{M.}},
\bauthor{\bsnm{Maurice}, \binits{L.}},
\bauthor{\bsnm{{Le Bourgeois}}, \binits{F.}},
\bauthor{\bsnm{Caseris}, \binits{M.}},
\bauthor{\bsnm{Gaschignard}, \binits{J.}},
\bauthor{\bsnm{Poline}, \binits{J.}},
\bauthor{\bsnm{Cohen}, \binits{R.}},
\bauthor{\bsnm{Titomanlio}, \binits{L.}},
\bauthor{\bsnm{Faye}, \binits{A.}},
\bauthor{\bsnm{Melki}, \binits{I.}},
\bauthor{\bsnm{Meinzer}, \binits{U.}}:
\batitle{Emergence of kawasaki disease related to sars-cov-2 infection in an
  epicentre of the french covid-19 epidemic: a time-series analysis}.
\bjtitle{The Lancet Child \& Adolescent Health}
\bvolume{4}(\bissue{9}),
\bfpage{662}--\blpage{668}
(\byear{2020}).
\doiurl{10.1016/S2352-4642(20)30175-9}
\end{barticle}
\endbibitem

\bibitem{angoulvant2020}
\begin{barticle}
\bauthor{\bsnm{Angoulvant}, \binits{F.}},
\bauthor{\bsnm{Ouldali}, \binits{N.}},
\bauthor{\bsnm{Yang}, \binits{D.D.}},
\bauthor{\bsnm{Filser}, \binits{M.}},
\bauthor{\bsnm{Gajdos}, \binits{V.}},
\bauthor{\bsnm{Rybak}, \binits{A.}},
\bauthor{\bsnm{Guedj}, \binits{R.}},
\bauthor{\bsnm{Soussan-Banini}, \binits{V.}},
\bauthor{\bsnm{Basmaci}, \binits{R.}},
\bauthor{\bsnm{Lefevre-Utile}, \binits{A.}},
\bauthor{\bsnm{Brun-Ney}, \binits{D.}},
\bauthor{\bsnm{Beaujouan}, \binits{L.}},
\bauthor{\bsnm{Skurnik}, \binits{D.}}:
\batitle{{Coronavirus Disease 2019 Pandemic: Impact Caused by School Closure
  and National Lockdown on Pediatric Visits and Admissions for Viral and
  Nonviral Infections—a Time Series Analysis}}.
\bjtitle{Clinical Infectious Diseases}
\bvolume{72}(\bissue{2}),
\bfpage{319}--\blpage{322}
(\byear{2020}).
\doiurl{10.1093/cid/ciaa710}
\end{barticle}
\endbibitem

\bibitem{vicuna2021}
\begin{barticle}
\bauthor{\bsnm{Vicuña}, \binits{M.I.}},
\bauthor{\bsnm{Vásquez}, \binits{C.}},
\bauthor{\bsnm{Quiroga}, \binits{B.F.}}:
\batitle{Forecasting the 2020 covid-19 epidemic: A multivariate quasi-poisson
  regression to model the evolution of new cases in chile}.
\bjtitle{Frontiers in Public Health}
\bvolume{9},
\bfpage{416}
(\byear{2021}).
\doiurl{10.3389/fpubh.2021.610479}
\end{barticle}
\endbibitem

\bibitem{endo2020}
\begin{barticle}
\bauthor{\bsnm{Endo}, \binits{A.}},
\bauthor{\bsnm{Abbott}, \binits{S.}},
\bauthor{\bsnm{Kucharski}, \binits{A.J.}},
\bauthor{\bsnm{Funk}, \binits{S.}}:
\batitle{Estimating the overdispersion in covid-19 transmission using outbreak
  sizes outside china}.
\bjtitle{Wellcome Open Research}
\bvolume{5},
\bfpage{67}
(\byear{2020}).
\doiurl{10.12688/wellcomeopenres.15842.3}
\end{barticle}
\endbibitem

\bibitem{VerHoef2007}
\begin{barticle}
\bauthor{\bsnm{Ver~Hoef}, \binits{J.M.}},
\bauthor{\bsnm{Boveng}, \binits{P.L.}}:
\batitle{Quasi-poisson vs. negative binomial regression: How should we model
  overdispersed count data?}
\bjtitle{Ecology}
\bvolume{88}(\bissue{11}),
\bfpage{2766}--\blpage{2772}
(\byear{2007}).
\doiurl{10.1890/07-0043.1}
\end{barticle}
\endbibitem

\end{thebibliography}


\end{document}